\documentclass[11pt]{article}
\usepackage{subcaption}

\usepackage{jheppub}

\usepackage{latexsym,cancel,amssymb}
\usepackage{graphicx}
\usepackage{epstopdf}   
\usepackage{epsfig}   
\usepackage{slashed, xcolor}
\usepackage{multirow}
\usepackage{verbatim}
\usepackage{booktabs}
\usepackage{hyperref}

\newcommand{\h}{{\rm h}}
\newcommand{\trh}{\tilde{\rm h}}

\newcommand{\beq}{\begin{equation}}
\newcommand{\eeq}{\end{equation}}
\newcommand{\bea}{\begin{eqnarray}}
\newcommand{\eea}{\end{eqnarray}}

\newcommand{\tTT}{{\tilde T}^{\rm TT}_{ij}}

\newcommand{\hij}{h_{ij}}

\newcommand{\D}{\mathrm{d}}
\newcommand{\gn}{G_\mathrm{N}}
\newcommand{\mpl}{M_\mathrm{Pl}}

\newcommand{\bk}{{\bf k}}
\newcommand{\bx}{{\bf x}}

\newcommand{\nn}{\nonumber \\}
\newcommand{\ogw}{\Omega_{\rm GW}}
\newcommand{\rgw}{\rho_{\rm GW}}
\newcommand{\rp}{\tilde{r}'}
\newcommand{\rD}{\tilde{r}}
\newcommand{\tD}{\tilde{t}}

\title{Gravitational waves and primordial black holes produced by dark meta stable vacuum decay}

\author[a,b]{Haipeng An}
\author[a]{Tingyu Li}
\author[a]{Chen Yang}
\affiliation[a]{Department of Physics, Tsinghua University, Beijing 100084, China}
\affiliation[b]{Center for High Energy Physics, Tsinghua University, Beijing 100084, China}

\emailAdd{anhp@mail.tsinghua.edu.cn}
\emailAdd{lity22@mails.tsinghua.edu.cn}
\emailAdd{yangc18@tsinghua.org.cn}

\abstract{

Inspired by string theory and cosmological constant problem, it is plausible that the Universe’s vacuum structure is characterized by a landscape of metastable vacua. The existence of dark matter and dark energy further suggests that the dark sector may inhabit its own “dark landscape.” If the dark vacuum is metastable, bubbles of lower-energy phases can nucleate at an approximately constant rate. Because the Hubble expansion rate is monotonically non-increasing with cosmic time, such nucleation can eventually lead to percolation and completion of a dark-sector phase transition. In this work, we investigate the phenomenological consequences of this transition, focusing on the resulting stochastic gravitational-wave background and the potential formation of primordial black holes. We find that the gravitational wave spectrum peaks at $k_{\mathrm{peak}} = 3.1 H_{\rm PT}$, with an amplitude $\Omega_{\mathrm{GW}}^{\mathrm{peak}} \simeq 0.3 \Omega_\gamma\, (\Delta \rho/\rho_{\rm tot})^2$. Furthermore, the formation of primordial black holes is suppressed due to $\Delta N_{\mathrm{eff}}$ constraint.

}

\begin{document}

\maketitle

\section{Introduction}

Various astrophysical observations indicate that the particles of the Standard Model (SM) account for only about five percent of the present-day energy density of the Universe; the remainder is attributed to dark matter (DM) and dark energy (DE). Yet the fundamental natures of DM and DE remain elusive.

The puzzle of why the observed DE density is so small lies at the heart of the cosmological constant (CC) problem, one of the most profound issues in theoretical physics. Several dynamical solutions have been proposed; however, to date the only broadly viable explanation is the anthropic argument~\cite{Weinberg:1987dv,Weinberg:1988cp}, which requires a landscape of vacua. The string landscape provides a vast set of low-energy vacua with different constants of nature, thereby making anthropic reasoning more concrete; conversely, anthropic reasoning renders the landscape potentially explanatory~\cite{Bousso:2000xa,Giddings:2001yu,Kachru:2003aw,Balasubramanian:2005zx,Susskind:2003kw,Polchinski:2006gy}. That said, constructing such a landscape ultimately demands a complete understanding of quantum gravity, which is still lacking.

DM and DE may reside in a sector—or in separate sectors—that interact with SM particles only via gravity, collectively referred to as the dark sector (DS). The total vacuum energy receives contributions from both the SM and the DS. It is therefore plausible that the DS possesses its own landscape of vacua and that, in the early Universe, it was trapped in a metastable vacuum. 

From the Friedmann equation, $H^2 \simeq \rho/(3\mpl^2)$, the Hubble rate $H$ decreases with time during standard cosmic evolution. Assuming no direct coupling between the DS and the SM thermal bath, the DS bubble nucleation rate per unit volume, $\Gamma/{\cal V}$, is approximately constant in time. The decay of the metastable vacuum then completes when $\Gamma/{\cal V} \times H^{-4} \gtrsim \mathcal{O}(1)$. In this paper we study the phenomenology of such a phase transition.

Throughout, we assume the DS has no nongravitational interactions with the SM. Consequently, the only observable imprints of the dark phase transition are gravitational waves (GWs), primordial black holes (PBHs), and dark radiation. In the remainder of this work we present detailed calculations of these signals.

During a first-order phase transition, the bubble nucleation rate per unit physical volume can be written as
\begin{align}
\frac{\Gamma}{\cal V} = m^4 e^{-S}\,,
\end{align}
where $m$ is the characteristic energy scale of the system. For vacuum transitions $S=S_4$, the 4D bounce action~\cite{Coleman:1977py,Callan:1977pt}; for thermal transitions $S=S_3/T$, with $S_3$ the 3D bounce action~\cite{Linde:1981zj}. In most models, the transition parameters evolve with the background. For example, in a thermal transition (see~\cite{Weir:2017wfa}), cooling reduces the barrier between the true and false vacua, thereby decreasing $S_3$ and increasing the nucleation rate; the transition typically completes when $\Gamma/{\cal V} \simeq H^4$. Another example is a first-order transition during inflation~\cite{An:2020fff,An:2022cce}, where the inflaton’s evolution triggers the transition. In both cases, the timescale is set by $\beta^{-1}$, where $\beta \equiv - dS/dt$. Usually $\beta \gg H$ is required for completion, so the transition finishes in a time $\beta^{-1} \ll H^{-1}$; in this regime, flat-spacetime simulations accurately capture bubble dynamics and GW production, and many numerical studies exist.

However,  in the DS scenario considered here (see Sec.~\ref{sec:meta}), the sector remains in a metastable vacuum until the Hubble rate falls below a critical value at which the transition can complete, leaving $H$ as the only relevant macroscopic scale. Consequently, cosmic expansion cannot be neglected when modeling bubble growth and collisions. To accurately compute the resulting GW power spectrum, we include the expansion of the Universe in lattice simulations (see Sec.~\ref{sec:GW}).

For transitions whose dynamics are driven by background evolution (thermal or inflation-triggered), the parameter $\beta$ sets the characteristic bubble size and hence the source quadrupole, leading to a GW amplitude suppressed by a factor $(H/\beta)^n$, with model-dependent $n$. By contrast, in our DS scenario, nucleation is approximately time-independent and the transition completes when $H$ becomes sufficiently small; as a result, there is no $(H/\beta)^n$ suppression of the GW signal strength (see Sec.~\ref{sec:GW}). 

In the DS phase transition considered here, the DS has no interactions with the SM beyond gravity. Consequently, the latent heat is deposited entirely within the DS as dark particles, and the scenario is constrained by limits on dark radiation. We therefore require $\Delta\rho/\rho_R \ll 1$, where $\Delta\rho$ is the latent energy density and $\rho_R$ is the total radiation energy density at the time the transition completes. A representative bound is $\Delta\rho/\rho_R \lesssim 0.14$.

Putting these elements together, the peak GW amplitude can be approximated as
\bea
\Omega_{\rm GW}^{\rm peak} \simeq C \,\Omega_\gamma\,\theta^2,
\eea
where $C$ is a numerical coefficient to be calibrated, $\Omega_\gamma = 5.38\times 10^{-5}$ is today’s photon energy-density fraction, and $\theta \equiv \Delta\rho/\rho_R$. In Sec.~\ref{sec:numerical}, numerical simulation gives $C = 0.3$. Numerical simulation also shows that the peak of the GW spectrum locates at $k_{\rm peak}=3.1 H$.

Forthcoming searches—including ground‑based detectors, space‑based interferometers, and pulsar timing arrays—aim for sensitivities to a stochastic background at the level $\Omega_{\rm GW} \sim 10^{-11}$–$10^{-15}$~\cite{KAGRA:2021kbb,ET:2025xjr,Evans:2021gyd,Luo:2025ewp,Ruan:2018tsw,Caprini:2009fx,Kawamura:2011zz,Reardon:2023gzh,EPTA:2015qep,NANOGrav:2023gor,Xu:2023wog,Moore:2014lga}. Since most of the latent energy is assumed to be converted into dark radiation, the bound $\Delta N_{\rm eff} < 0.4$~\cite{DESI:2024mwx,Allali:2024cji} implies that, absent significant dark‑sector cooling, $\theta \lesssim 0.1$. Accordingly, in this work we focus on the range $10^{-5} \lesssim \theta \lesssim 0.1$.

In generic regions of parameter space (without severe fine‑tuning or extremely small couplings), one expects $\Delta\rho \sim m^4$. During radiation domination, $H \sim T^2/\mpl$. Neglecting order‑one factors (including mild $g_*$ dependence), this gives
\bea
m \sim \theta^{1/4}\,T \gg H.
\eea
We will use this inequality throughout when deriving properties of the GW and PBH signals.

As in supercooled phase transitions~\cite{Sato:1981gv,Kodama:1982sf,Blau:1986cw,Liu:2021svg,Baker:2021sno,Hashino:2021qoq,He:2022amv,Hashino:2022tcs,Kawana:2022olo,Lewicki:2023ioy,Gouttenoire:2023naa,Gouttenoire:2023bqy,Salvio:2023ynn,Baldes:2023rqv,Salvio:2023blb,Conaci:2024tlc,Lewicki:2024ghw,Flores:2024lng,Kanemura:2024pae,Cai:2024nln,Goncalves:2024vkj,Banerjee:2024cwv,Wu:2024lrp,Hashino:2025fse,Ghoshal:2025dmi,Kierkla:2025vwp}, false-vacuum remnants can form during DS metastable-vacuum decay. The density contrast of these remnants relative to the surrounding Universe grows with expansion, and they may ultimately collapse into black holes. As shown in Sec.~\ref{sec:PBH}, because the nucleation rate in our scenario is approximately time independent, the number density of PBHs can be exponentially enhanced compared to thermal phase transitions.

We further predict a correction proportional to $k^3\log^2(k/H)$ to the infrared (IR) tail of the spectrum, which originates from the long-term post-collision evolution of the bubble walls.
On the other hand, the predicted number density of primordial black holes (PBHs) is suppressed, due to the constraint on the dark sector energy density imposed by $\Delta N_{\rm eff}$.

Gravitational waves (GWs) from phase transitions in DS with a temperature independent of the Standard Model have been considered in the literature (see, e.g., \cite{Fairbairn:2019xog}). However, the characteristic features of the resulting GW spectrum have not yet been examined in detail. For supercooled phase transitions, GW production computed within the envelope approximation was pioneered in \cite{Yamada:2025hfs,Yamada:2025cfr}. As shown in Sec.~\ref{sec:GW}, however, in DS metastable vacuum decay the bubble walls can become highly boosted, and the GW source can persist for a duration far exceeding a Hubble time. Consequently, the envelope approximation is not suitable for capturing the qualitative features of the GW spectrum in this regime.

The rest of the paper is organized as follows. In Sec.~\ref{sec:meta} we discuss the decay of the metastable DS vacuum during the radiation-dominated era and present formulas for the decay rate and the fraction converted to the new vacuum. In Sec.~\ref{sec:GW} we present our analysis of the GW signal. In Sec.~\ref{sec:numerical} we numerically simulate the phase transition and GW signal.   In Sec.~\ref{sec:PBH} we compute the PBH abundance induced by the DS phase transition. We conclude in Sec.~\ref{sec:summary}.

\section{Meta Stable Vacuum Evolution and Bubble Dynamics}
\label{sec:meta}

In this section, we analyze the false-vacuum decay process and the corresponding bubble dynamics. While there are similarities to thermal first-order phase transitions, there are two essential differences:

\begin{itemize}
    \item Time dependence of nucleation: In contrast to thermal first-order phase transitions, where the bounce action depends on temperature and the nucleation rate increases as the Universe cools, the decay rate per unit physical volume is approximately time-independent in this scenario. Consequently, the nucleation-rate density does not increase with time.
    
    \item Energy deposition and wall dynamics: In the absence of a coupled fluid or plasma, the released vacuum energy is channeled into the bubble walls, which then undergo runaway acceleration and reach extremely large Lorentz factors. Consequently, the field’s kinetic energy far exceeds its potential energy, and this hierarchy persists even during and after collisions. The total energy remains kinetic dominated, so the post-collision field configuration retains a bubble‑wall-like profile.
\end{itemize}

Using numerical results, we qualitatively examine bubble-wall evolution, including nucleation, expansion, and pre- and post-collision dynamics. The numerical setup is described in Sec.~\ref{sec:numerical}.

\subsection{Bubble nucleation and expansion before collision}
\label{subsec:before_collision}

For concreteness, we model the DS vacuum decay with a scalar field $ \phi $. We assume that, in the early Universe, the DS underwent a first-order phase transition during which $\phi$ tunneled from a local minimum $\phi_F$ to another local minimum $\phi_T$, with $V(\phi_T) < V(\phi_F)$. The tunneling rate from $\phi_F$ to $\phi_T$ is governed by the bounce action~\cite{Callan:1977pt,Coleman:1977py}. Immediately after nucleation, the bubble radius is order of $m^{-1}$, much smaller than $ H^{-1} $, where $m$ is the characteristic DS mass scale. Hence the Hubble expansion can be neglected during nucleation. The decay rate per unit physical volume scales as
\begin{align}
    \frac{\Gamma}{\cal V}\sim m^4 e^{-S_4} \ ,
\end{align}
where $S_4$ is the action of the $SO(4)$-symmetric bounce, satisfying
\begin{align}\label{eq: bounce Eom}
    \frac{\mathrm{d}^2\phi}{\mathrm{d}s^2}+\frac{3}{s}\frac{\mathrm{d}\phi}{\mathrm{d}s}=\frac{\mathrm{d}V}{\mathrm{d}\phi}\ ,
\end{align}
with boundary conditions
\begin{align}
\phi(\infty)=\phi_F,\qquad\left.\frac{\mathrm{d}\phi}{\mathrm{d}s}\right|_{s=0}=0 \ ,
\end{align}
and $s \equiv \rho = \sqrt{|\mathbf{x}|^2 + \tau^2}$ is the Euclidean radial coordinate (with Euclidean time $\tau$).

After a bubble nucleates, its wall is driven outward by the pressure difference between the two vacua and expands until it collides with other bubbles. During this stage, the expansion of the Universe must be taken into account. We work with the spatially flat Robertson–Walker metric,
\begin{align}\label{eq: frw metric}
    \mathrm{d}s^2 = \mathrm{d}t^2 - a^2\,\mathrm{d}\mathbf{x}^2
= a^2\,\big(\mathrm{d}\eta^2 - \mathrm{d}\mathbf{x}^2\big)\ ,
\end{align}
where $a$ is the scale factor, and $t$ and $\eta$ denote physical and conformal time, respectively. We focus on phase transitions during radiation domination, for which
\begin{align}
    a = \left(\frac{t}{t_{\mathrm{PT}}}\right)^{1/2}, \qquad \eta = 2(t_{\mathrm{PT}} t)^{1/2} \ , \qquad a = H_{\mathrm{PT}}\,\eta, \qquad H = \frac{1}{2t} \ ,
\end{align}
with $H_{\mathrm{PT}} \equiv H|_{a=1}$. For later convenience, we define $t_{\mathrm{PT}}$
as the time at which, on average, one bubble is nucleated per Hubble volume per Hubble time; equivalently,
\begin{align}
\frac{\Gamma}{{\cal V}} = H_{\mathrm{PT}}^4 = \left(\frac{1}{2t_{\mathrm{PT}}}\right)^4 \ ,
\end{align}
and $\eta_{\mathrm{PT}}$ as the conformal time when $t=t_{\mathrm{PT}}$.

In what follows, we use whichever time coordinate is most convenient: conformal time $\eta$ for gravitational-wave production and evolution, and the physical time $t$ for discussing PBH formation.

For a spherically symmetric configuration, the scalar EoM is
\begin{align}\label{eq: bubble EoM}
    \frac{\partial^2\phi}{\partial\eta^2}+2\mathcal{H}\frac{\partial\phi}{\partial \eta}-\frac{1}{r}\frac{\partial^2}{\partial r^2}\left(r\phi\right)+a^2\frac{\mathrm{d}V}{\mathrm{d}\phi}=0 \ ,
\end{align}
where $ \mathcal H \equiv a^{-1} \mathrm da/\mathrm d\eta = 1/\eta $ is the conformal Hubble parameter and $ r \equiv |\mathbf{x}| $ is the comoving radius.

When the bubble’s curvature radius is much larger than $ m^{-1} $, the physical wall thickness is set by microphysics and the boost factor. The bubble profile can then be described by the ansatz~\cite{Cai:2020djd}:
\begin{align}\label{eq:bubble wall ansatz}
    \phi(\eta,r)=\phi_s[a\gamma(\eta)(r-r_s(\eta))] \ ,
\end{align}
where $ \phi_s $ is the wall profile in its rest frame, $ r_s(\eta) $ is the comoving wall position, $ v_w \equiv \partial_\eta r_s $ is the comoving wall velocity, and $ \gamma = 1/\sqrt{1 - v_w^2} $ is the corresponding boost factor. In the absence of the Hubble expansion, the wall approaches the speed of light; as the bubble radius approaches the Hubble scale, Hubble friction becomes important, which is encoded in the evolution of $ r_s(\eta) $.

Substituting the ansatz Eq.~\eqref{eq:bubble wall ansatz} into the field equation yields an effective EoM for the wall~\cite{Lewicki:2023ioy,Cai:2020djd}:
\begin{align}\label{eq: bubble eff eom}
    \partial_\eta\left(\gamma v_w\right)+3\mathcal{H} \gamma v_w +\frac{2\gamma}{r_s}=a\frac{\delta V}{\sigma} \ ,
\end{align}
where $ \delta V \equiv V(\phi_F) - V(\phi_T) $ is the vacuum energy difference and
\begin{align}
    \sigma = \int_{-\infty}^{\infty}\left(\frac{\mathrm{d}\phi_s}{\mathrm{d}\tilde{r}}\right)^2\mathrm{d}\tilde{r} \ ,
\end{align}
is the wall tension with $\tilde{r}=a\gamma(\eta)(r-r_s(\eta))$. The detailed derivation of Eq.~\eqref{eq: bubble eff eom} is presented in Appendix~\ref{sec:AppendixA}. 

We define the conformal time of bubble nucleation as $\eta_0$.
At nucleation, the critical configuration is static such that $ v_{w} (\eta_0)= 0 $ and $\gamma(\eta_0) = 1$. Then Substituting into the effective equation~\eqref{eq: bubble eff eom} gives
\begin{align}
    r_{cr}=\frac{2\sigma}{a\delta V} \ ,
\end{align}
where $r_{cr}$ is the comoving critical bubble radius. Therefore, the physical critical radius is $R_{cr} = 2\sigma/\delta V\sim m^{-1}$ as expected by naive dimensional analysis. 
%We assume the Universe is in RD era, then $a=H_0\eta$. Here $H_0$ is the Hubble parameter at $a=1$. 
As we will show below, bubbles typically reach radii of or order $H^{-1}$ by the time they collide. Their walls therefore accelerate to ultra-relativistic speeds, implying $v_w \approx 1$.

Then we can solve Eq.~\eqref{eq: bubble eff eom}. $v_w\approx 1$ implies that $r_s\approx \eta-\eta_0$, and $\mathcal{H}=1/\eta$. 
The solution of Eq.~\eqref{eq: bubble eff eom} yields
\begin{align}\label{eq: solution before collision}
    \gamma(\eta)\sim a(\eta)(\eta-\eta_0)\frac{\delta V}{\sigma} \sim \frac{\eta-\eta_0}{ r_{cr} }\ .
\end{align}
For a phase transition whose characteristic energy scale is much smaller than the Planck scale yet contributes a sizable fraction of the Universe's energy density, the comoving critical radius satisfies $r_{cr} \ll H^{-1}$, since $H$ is Planck-scale suppressed. 
Therefore $\gamma \gg 1$ for most of the pre-collision expansion. 
%Therefore, in most realistic scenarios, we have $\gamma_{\rm wall} \gg 1$.  The bubble walls thus behave like radiation component before collision. 
%

\subsection{Evolution of field configuration after bubble wall collisions}
\label{subsec:after_collision}

Since the boost factor $\gamma \gg 1$, the kinetic terms (time- and spatial-derivative terms) in Eq.~\eqref{eq: bubble EoM} dominate over the force term $\partial V/\partial \phi$. Consequently, immediately after the collision the field retains a wall-like profile, with most of the energy remaining localized in the thin shell. Behind the wall, however, the field is no longer exactly at the true vacuum; it begins to oscillate and generates a wave-like tail trailing the leading shell.

To analyze the post-collision evolution of the energy shells, we identify three relevant length scales: the Hubble length $H^{-1}$, the microscopic scale set by the mass parameter $m^{-1}$, and the wall thickness at the time of collision $(\gamma m)^{-1}$. They satisfy
\bea\label{eq:scale relation}
H^{-1} \gg m^{-1} \gg (\gamma m)^{-1}\,.
\eea
To isolate the large-boost dynamics of the energy shell, we begin by neglecting cosmic expansion and studying the evolution in Minkowski spacetime. We then generalize these results to the expanding universe.

\subsubsection{Evolution of energy shells in Minkowski spacetime}
\label{sec: flat bubble case}

Here since we have neglected the expansion of the Universe, the coordinate system and metric we use is 
\bea
\D s^2 = \D t^2 - \D x^2-\D y^2-\D z^2 \ .
\eea
Also, when we neglect the expansion of the Universe, the largest length scale $H^{-1}$ is now replaced by $r_c$, which is the bubble radius at collision the time $t_c$. Then Eq.~\eqref{eq:scale relation} is replaced by
\begin{align}
    r_c \gg m^{-1} \gg (\gamma m)^{-1}\ ,
\end{align}
and the relation $\gamma\sim m/H$ is replaced by $\gamma \sim m r_c$. 

\begin{figure}[htpb]
\centering
\includegraphics[width=0.8\linewidth]{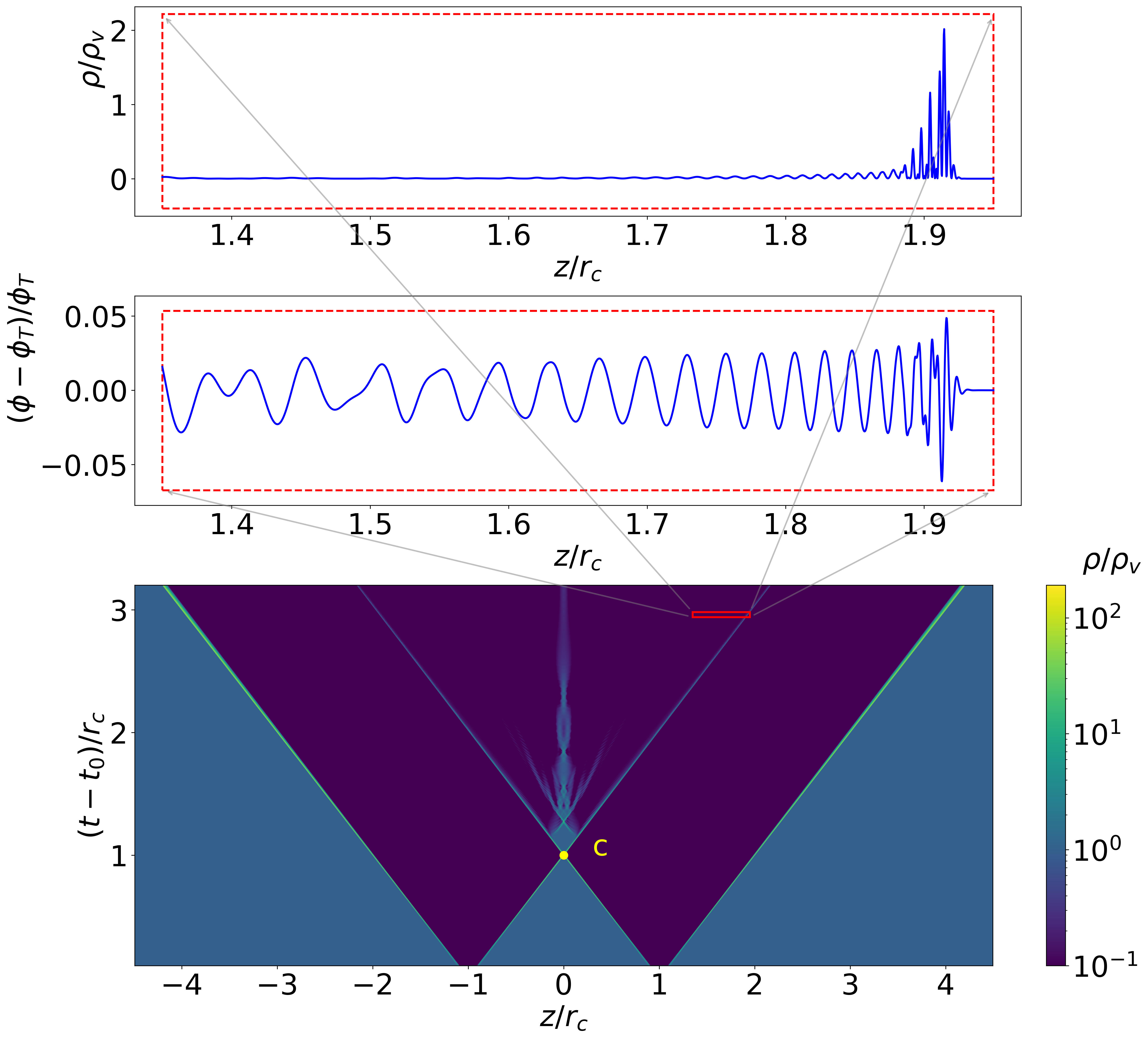}
\caption{The lower panel displays the energy density distribution before and after the collision of two bubbles at point $c$. The upper and middle panels zoom in on the region outlined by the red rectangle, showing the detailed density profile and field configuration of a segment of the bubble wall, respectively. For this specific collision event, the boost factor is $\gamma = 10$. We observe that despite the development of multiple oscillations in the post-collision waveform, the majority of the energy remains concentrated in a thin region near the propagating wavefront of the bubble shell. }
\label{fig:bubble collision}
\end{figure}

To illustrate this process and highlight the importance of the large boost, Fig.~\ref{fig:bubble collision} shows the numerical evolution of the field configuration after the collision of two bubbles, computed while neglecting cosmic expansion. The lower panel displays the evolution of the field energy density distribution. The two bubbles nucleate at $t_0$ and start to expand, and collide at the time $t_c$. The boost of bubble walls are set to be 10 when they collide with each other.\footnote{In realistic scenarios the typical boost factor at collision can reach $\mpl/m$.} The colors indicate the energy density normalized to the false vacuum energy density. The upper and middle panels show, respectively, the energy density distribution and field configuration on a finite equal-time slice, indicated by the red rectangular box. One clearly sees that after the collision the energy density of the field remains localized in a very thin shell that propagates at nearly the speed of light, so the kinetic energy continues to dominate the shell’s energy. 

From Fig.~\ref{fig:bubble collision}, we also see that after the collision, even in the flat spacetime approximation, the amplitude of the field configuration decays with time. Consequently, the nonlinear terms in $\partial V/\partial\phi$ becomes less important than the mass term. When analyzing the propagation of the energy shell, we can therefore neglect the nonlinear terms and retain only the mass term, so that the EoM for shell reduces to
\begin{align}\label{eq:flatEoM}
\partial^2_t \phi - \nabla^2\phi + m^2 \phi = 0 \ ,
\end{align}
which is simply the EoM of a massive scalar field. 

To study the evolution of the energy shells after the bubble collision, it is convenient to use the spherical coordinates and decompose $\phi$ in spherical harmonics. Eq~\eqref{eq:flatEoM} then becomes 
\begin{equation}\label{eq:flatEoM2}
\partial_t^2 \chi_{\ell m} -\partial_r^2 \chi_{\ell m} + \frac{\ell(\ell+1)}{r^2} \chi_{\ell m} + m^2 \chi_{\ell m} = 0 \ ,
\end{equation}
where $\chi_{\ell m} = r \phi_{\ell m}$, and  $\phi_{\ell m}$ are the spherical-harmonic components of $\phi$. After the collision, the shells remain highly boosted, with $\gamma \gg 1$. The first two terms in Eq.~\eqref{eq:flatEoM2} are ${\cal O}(\gamma^2)$, whereas the last two terms are ${\cal O}(\gamma^0)$. Moreover, using the relation that $\gamma \sim m r_c$, we have $r^{-2}\sim m^2/\gamma^2\ll m^2$, so the third term can be neglected. As a result, even after the collision, the evolution of the shells is still governed by the spherical EoM. 

It is important to note that, although the EoM is spherical, different patches of the spherical shell can have different boost factor $\gamma$, since different parts of a bubble typically collide with other bubbles at different times and thus experience different acceleration histories. It is precisely these variations in $\gamma$ on the nearly spherical shell that source GWs after collisions. 

\begin{figure}[htpb]
\centering
\includegraphics[width=0.98\linewidth]{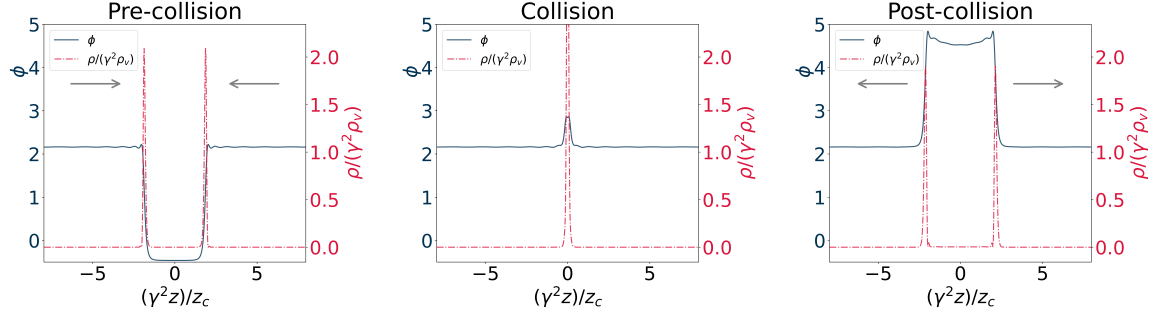}
\caption{Bubble collision process in three sequential stages. a) Pre-collision: The left panel depicts a kink-like configuration with false vacuum between the two bubbles. b) Collision: The middle panel shows the moment the bubble walls merge. c) Post-collision: The right panel displays the outcome, where most of the energy is carried by two outward-propagating shells. (The simulation uses a boost factor of $\gamma=80$.)}
\label{fig:after bubble collision}
\end{figure}

After the collision, each $\ell$-mode evolves in almost the same way, with differences suppressed by $\gamma^{-2}$. The EoM can therefore be further simplified to 
\bea\label{eq:flatEoM3}
\partial_t^2 \chi_{\ell m} - \partial_r^2 \chi_{\ell m} + m^2 \chi_{\ell m} = 0 \ ,
\eea
which we use to study the oscillatory behavior and energy distribution shown in Fig.~\ref{fig:bubble collision}. Thus, $\chi_{\ell m}$ satisfies the 1+1D Klein-Gordon equation~\eqref{eq:flatEoM3}, which can be solved using the Green's function method. Suppose the field satisfies the following initial condition at the collision time $t_c$:
\begin{align}\label{eq: initial c pulse}
    \chi_{\ell m}(\rD+r_c,t_c)=\phi_c(\rD),\qquad \partial_t \chi_{\ell m}(\rD+r_c,t_c)=\psi_c(\rD)\ .
\end{align}
For convenience, we introduce the shifted coordinates $ \rD=r-r_c$ and $\tD=t-t_c$, and we express the initial conditions in terms of these variables. The solution can be then written as 
\begin{align}\label{eq: greens full}
    \chi_{\ell m}(r,t)= & \frac{1}{2}[\phi_c(\rD-\tD)+\phi_c(\rD+\tD)] \nonumber\\
& -\frac{m \tD}{2} \int_{\rD-\tD}^{\rD+\tD} \frac{J_1\left(m \sqrt{\tD^2-\left(\rD-\rp\right)^2}\right)}{\sqrt{\tD^2-\left(\rD-\rp\right)^2}} \phi_c\left(\rp\right) d \rp \nonumber\\
& +\frac{1}{2} \int_{\rD-\tD}^{\rD+\tD} J_0\left(m \sqrt{\tD^2-\left(\rD-\rp\right)^2}\right) \psi_c\left(\rp\right) d \rp\ ,
\end{align}
We have used the Green's function for the massive Klein-Gordon equation. During the collision, the two bubble walls form a delta-function-like configuration that subsequently splits into two shells propagating in opposite directions, as illustrated in Fig.~\ref{fig:after bubble collision}. The wavefronts of the two shells carry most of the system's energy. In what follows, We focus on the evolution of the right-moving shell, which propagates in the positive $r$ direction and generates the waveform shown in Fig.~\ref{fig:bubble collision}. Accordingly, we take the right-moving shell immediately after collision as the initial condition for the EoM.

After the collision, the shell carries most of the pre-collision energy, and its boost factor $\gamma \sim m r_c$ is likewise inherited from the pre-collision state. Consequently, the kinetic energy of the bubble shell dominates over the potential energy. This dominance implies that, immediately after the collision, $\chi_{\ell m}(r,t)$ at $t = t_c$ approximately satisfies the EoM $\partial_t^2\chi - \partial_r^2\chi = 0$. This equation admits right-moving solutions of the form $\chi_{\ell m}(r,t) \sim \chi_{\ell m}(t-r)$ near $t = t_c$. Hence, the two sets of initial data are related by  
\begin{align}\label{eq:initial relation}
    \psi_c(\rD)
    = \partial_t \chi_{\ell m}(\rD + r_c - t)\big|_{t = t_c}
    = -\partial_{\rD} \phi_c(\rD)\,.
\end{align}
The profile $\phi_c$ is localized within a region of width $1/(\gamma m)$ around $\Delta r = 0$. Therefore, $\phi_c(r)$ is nonzero only for $0 \lesssim r \lesssim 1/(\gamma m)$.

Considering only the shell moving to the right-hand side, we can neglect the $\phi_c(\rD+\tD)$ term in Eq.~\eqref{eq: greens full}. Using Eq.~\eqref{eq:initial relation} and performing an integration by parts, we obtain
\begin{align}
    \chi_{\ell m}(r,t)&= \phi_c(\rD-\tD)-\frac{m(\tD+\rD)}{2}\int_{\rD-\tD}^{\rD+\tD} \frac{J_1\left[m s(\rD')\right]}{s(\rD')} \phi_c\left(\rp\right) d \rp \label{eq:first form}\ ,\\
& =\frac{1}{2}\left(1-\frac{\tD}{\rD}\right)\phi_c(\rD-\tD)\nonumber\\
&\hspace{2cm}-\frac{1}{2} \int_{\rD-\tD}^{\rD+\tD} J_0\left[m s(\rp)\right] \left(1+\frac{\tD}{\rD}\right)\partial_{\rp}\phi_c\left(\rp\right) d \rp\label{eq:second form}\ ,
\end{align}
where $s(\rp)=\sqrt{\tD^2-\left(\rD-\rp\right)^2}\approx \sqrt{(\tD+\rD)(\tD-\rD+\rp)}$. Here we have used the condition $\rD, \tD \gg m^{-1} \gg \rD'$, since $\phi_c(r)$ is nonzero only within the region $r\sim (\gamma m)^{-1}$. 

In the following, we study the qualitative behavior of $\chi_{\ell m}$. We assume that the shell propagates with velocity $v$, with Lorentz factor $\gamma = (1 - v^2)^{-1/2}$. As shown in Fig.~\ref{fig:bubble collision}, the shell profile also evolves during propagation. We define $v$ as the velocity of the peak of the energy density. To analyze the qualitative features of the shell configuration, we divide the profile into two regions according to the peak position: $\rD < v \tD$ and $\rD > v \tD$. In both regions, the waveform decays as a function of $|\rD - v \tD|$.

As we will see in Sec.~\ref{sec:GW}, the discussion below will be used to derive the infrared part of the GW spectrum, which is more sensitive to the long-term evolution of the shells. We therefore focus on the regime $\rD, \tD > r_c$. For the region $\rD, \tD \lesssim r_c$, although an analytic treatment is not available, numerical simulations show that the energy density evolves in the same qualitative manner.

\begin{itemize}

    \item $\rD < v \tD$
    %\item $\tD-\rD\gtrsim (4\gamma^2)^{-1}(\tD+\rD)$. 
    
    In this region, together with the condition $\gamma \gg 1$, we can show that  
    \begin{align}
    \tD - \rD > \frac{1}{4\gamma^2}\,(\tD + \rD) \,.
    \end{align}
    Combining this with $\rD, \tD > r_c$, we see that the first term in Eq.~\eqref{eq:first form}, $\phi_c(\rD-\tD)$, vanishes, since $\phi_c(r)$ is nonzero only for $0 \lesssim r \lesssim 1/(\gamma m)$.  

    In the $\rp$ integral, $s(\rp)$ can be approximated as  
    \begin{align}\label{eq:expands}
    s(\rD') = \sqrt{(\tD+\rD)(\tD-\rD+\rp)}
    \approx \sqrt{\tD^2-\rD^2}
    + \frac{1}{2}\sqrt{\frac{\tD+\rD}{\tD-\rD}}\,\rp \,.
    \end{align}
    In the same regime, we have  
    \begin{align}
    \left|\frac{1}{2}\sqrt{\frac{\tD+\rD}{\tD-\rD}}\,\rp\right|
    < \frac{1}{2} \cdot 2\gamma \cdot \frac{1}{\gamma m} \sim m^{-1} \,.
    \end{align}
    Therefore, in Eq.~\eqref{eq:first form}, we may approximate $s(\rp) \approx s(0) \equiv s$ when evaluating the $\rp$ integral. In this region, we thus obtain the analytic approximation  
    \begin{align}
    \chi_{\ell m}(r,t) \approx -\frac{\bar{\phi}_c}{2r_c}\left(\tD+\rD\right)\frac{J_1\!\left(m s\right)}{m s}\ ,
    \end{align}
    where  
    \begin{align}
    \bar{\phi}_c \equiv \frac{1}{m \gamma} \int_{\rD - \tD}^{\rD + \tD} \phi_c(\rp)\, d \rp
    \end{align}
    can be interpreted as the average value of $\phi_c$ over its support.

    \item $\rD > v\tD$
    
   In this regime, we have  
    \bea
    v^{-1} = \frac{\tD}{\rD} < 1 + \frac{1}{2\gamma} \,.
    \eea
    Thus, the contribution of the first term in Eq.~\eqref{eq:second form} to $\chi_{\ell m}$ is suppressed by a factor of $\gamma^{-2}$ and can be neglected. Moreover, we may set $\tD = \rD$ in the integral in Eq.~\eqref{eq:second form}, obtaining
    \begin{align}\label{eq:third form}
    \chi_{\ell m}(r,t)
    = -\int_{\rD-\tD}^{\rD+\tD}
    J_0\!\left[m\,s(\rp)\right]\,
    \partial_{\rp}\phi_c\!\left(\rp\right)\, d \rp\,.
    \end{align}

    This region lies ahead of the main waveform and beyond the energy-density peak of the shell. Consequently, the field amplitude decreases rapidly with $\rD$, as illustrated in the second panel of Fig.~\ref{fig:bubble collision}.

    Our goal here is to qualitatively describe the oscillatory behavior observed in this region (again, see the second panel of Fig.~\ref{fig:bubble collision}). From Eq.~\eqref{eq:third form}, we see that oscillations appear where $m s > 1$. Focusing on this domain and recalling that $s = (\tD^2 - \rD^2)^{1/2}$, the condition $m s > 1$ implies
    \bea
    \tD - \rD
    > m^{-2}(\tD + \rD)^{-1}
    = \frac{r_c^2}{\gamma^2 (\tD + \rD)}\,.
    \eea

    Since $s(\rp) \geq s$, we have $m s(\rp) > 1$ throughout the integration range. We can therefore use the asymptotic form of $J_0$ in Eq.~\eqref{eq:third form} and obtain
    \begin{align}\label{eq: expand chi small r}
    \chi_{\ell m}(r,t)
    &\approx -\sqrt{\frac{2}{\pi}}
    \int_{\rD-\tD}^{\rD+\tD}
    \frac{\cos\!\left[m s(\rp) - \frac{\pi}{4}\right]}
         {\sqrt{m s(\rp)}}\,
    \partial_{\rp}\phi_c\!\left(\rp\right)\, d \rp \nonumber\\[4pt]
    &\approx -\sqrt{\frac{2}{\pi}}\,
    \frac{\cos\!\left(m s - \frac{\pi}{4}\right)}{\sqrt{m s}}
    \int_{\rD-\tD}^{\rD+\tD}
    \cos(k_{\rm eff} \rp)\,
    \partial_{\rp}\phi_c\!\left(\rp\right)\, d \rp \nonumber\\[4pt]
    &\quad - \sqrt{\frac{2}{\pi}}\,
    \frac{\sin\!\left(m s - \frac{\pi}{4}\right)}{\sqrt{m s}}
    \int_{\rD-\tD}^{\rD+\tD}
    \sin(k_{\rm eff} \rp)\,
    \partial_{\rp}\phi_c\!\left(\rp\right)\, d \rp\,,
    \end{align}
    where in the second step we used the expansion~\eqref{eq:expands}, defined the effective momentum
    \bea
    k_{\rm eff} = \frac{m}{2}\sqrt{\frac{\tD+\rD}{\tD-\rD}}\,,
    \eea
    and neglected the $\rp$-dependence in the denominator.

    Remarkably, the integrals in Eq.~\eqref{eq: expand chi small r} have the structure of Fourier transforms of $\partial_r \phi_c$, with an effective frequency $k_{\rm eff}$ that depends on $\tD - \rD$. Since we are considering the region $\rD > v \tD$, and therefore $\tD-\rD < (4\gamma^2)^{-1}(\tD+\rD)$, we have $k_{\rm eff} > m\gamma$. The initial profile $\partial_r \phi_c$ is characterized by a momentum cutoff of order $m\gamma$, so its Fourier components with $k > m\gamma$ are typically exponentially suppressed. Hence, we expect the integrals in Eq.~\eqref{eq: expand chi small r} to behave as
    \begin{align}
    \int_{\rD-\tD}^{\rD+\tD}
    \cos(k_{\rm eff} \rp)\,
    \partial_{\rp}\phi_c\!\left(\rp\right)\, d \rp
    &\sim
    \int_{\rD-\tD}^{\rD+\tD}
    \sin(k_{\rm eff} \rp)\,
    \partial_{\rp}\phi_c\!\left(\rp\right)\, d \rp \nonumber\\
    &\sim
    \exp\!\left[-\left(\frac{k_{\rm eff}}{m\gamma}\right)^{\alpha}\right]\bar{\phi}_c\,,
    \end{align}
    where $\alpha$ depends on the detailed shape of the initial profile and is treated as a model-dependent parameter.

    Substituting this estimate into Eq.~\eqref{eq: expand chi small r}, we obtain
    \begin{align}
    \chi_{\ell m}(r,t)
    &\sim -\sqrt{\frac{2}{\pi}}\,
    \frac{\cos\!\left[m s + f(k_{\rm eff})\right]}{\sqrt{m s}}\,
    \bar{\phi}_c\,
    \exp\!\left[-\left(\frac{k_{\rm eff}}{m\gamma}\right)^{\alpha}\right] \nonumber\\[4pt]
    &\sim -\sqrt{\frac{2}{\pi}}\,
    \frac{\cos\!\left[m s + f(k_{\rm eff})\right]}{\sqrt{m s}}\,
    \bar{\phi}_c\,
    \exp\!\left[-\left(\frac{1}{2\gamma}
    \sqrt{\frac{\tD+\rD}{\tD-\rD}}\right)^{\alpha}\right],
    \end{align}
    where $f(k_{\rm eff})$ is an effective phase. Although this is not the exact solution for $\chi_{\ell m}$ in the regime $\tD-\rD\lesssim (4\gamma^2)^{-1}(\tD+\rD)$, it nonetheless captures essential qualitative features that will be useful later.

\end{itemize}

In summary, We have derived of the qualitative behavior of $\chi_{\ell m}$. Using the definition $\chi_{\ell m}=r\phi_{\ell m}$, we obtain the following expressions for the spherical-harmonic modes $\phi_{\ell m}$ of the field $\phi$. In the region $\rD > v \tD$ (equivalently $\tD-\rD > {(\tD+\rD)}/({4\gamma^2})$), 
\begin{align}\label{eq:phiL}
    \phi_{\ell m}(r,t)= -\frac{\bar{\phi}_c}{2rr_c}\left(\tD+\rD\right)\frac{J_1\left(m s\right)}{ms} \approx - \sqrt{\frac{2}{\pi}} \frac{\cos\left[ ms - \frac{3\pi}{4} \right]}{r\sqrt{ms}} \bar\phi_c  \frac{1}{2\gamma} \sqrt{\frac{\tD + \rD}{\tD - \rD}} \ ,
\end{align}
where we have used $ms > 1$ and $\gamma \gg 1$. For $\rD > v \tD$ ($\tD-\rD < ({\tD+\rD})/({4\gamma^2})$),
\begin{align}\label{eq:phiR}
    \phi_{\ell m}(r,t)\sim -\sqrt{\frac{2}{\pi}}\frac{\cos\left[ms+f(k_{eff})\right]}{r\sqrt{ms}}\bar{\phi}_c \exp\left[-\left(\frac{1}{2\gamma}\sqrt{\frac{\tD+\rD}{\tD-\rD}}\right)^\alpha\right]\ .
\end{align}

With these qualitative solutions in hand, we are now ready to discuss their implications in the following.

\begin{itemize}
    \item The peak amplitude of $\phi_{\ell m}$

From Eqs.~\eqref{eq:phiR} and \eqref{eq:phiL}, we observe that, in the region $\tD > v \rD$, the amplitudes of the $\phi$ oscillation decays with $\tD - \rD$, whereas in the region $\tD < v \rD$, the amplitudes decays exponentially. They both approaches the peak value 
\bea
\phi_{\ell m}^{\rm peak} \sim \frac{\bar\phi_c}{r} \sqrt{\frac{r_c}{\tD + \rD}} \ .
\eea
We can see that for $\rD,\tD \gg t_c$, the peak of field strength scales as $r^{-3/2}$. 

\item Gradient energy density distribution

The gradient energy of the shell is crucial for computing the GW production. It is defined as  
\bea
\rho_{g}(r,t)=\frac{1}{2}\left(\frac{\mathrm{d}\phi}{\mathrm{d}r}\right)^2 .
\eea
Using Eqs.~\eqref{eq:phiL} and~\eqref{eq:phiR}, we obtain, for $\,\rD < v \tD$,  
\begin{align}\label{eq: left half integration}
\rho_g(r,t)
&\approx \frac{\bar{\phi}_c^{\,2}(\tD+\rD)^2 \rD^2}{8\pi r^2 r_c^2}\,\frac{1}{m s^{5}}
\approx \frac{m\,\bar{\phi}_c^{\,2}}{32\pi r^2 \gamma^2}\,
\frac{(\tD+\rD)^{3/2}}{(\tD-\rD)^{5/2}} \, .
\end{align}
Here we have used the asymptotic form of the Bessel function as in Eq.~\eqref{eq:phiL}, kept only the leading contribution from spatial derivatives, and averaged over rapid oscillations via $\cos^2(ms)\to 1/2$. We also used $\,\tD+\rD \approx 2 \rD \approx 2 \tD$ in this regime.

For $\,\rD > v \tD$, using the same approximations, $\rho_g$ can be written as  
\bea\label{eq: right half integration}
\rho_g(r,t)\sim \frac{m \bar{\phi}_c^{\,2}}{8\pi r^2}\,
\frac{(\tD + \rD)^{1/2}}{(\tD - \rD)^{3/2}}\,
\exp\!\left[-2\left(\frac{1}{2\gamma}
\sqrt{\frac{\tD + \rD}{\tD - \rD}}\right)^{\alpha}\right].
\eea

Introducing the variable
\bea\label{eq:u}
u \equiv 2\gamma \sqrt{\frac{\tD - \rD}{\tD + \rD}} \,,
\eea
$\rho_g(r,t)$ can be recast as  
\bea\label{eq:rho g}
\rho_g(r,t)=
\begin{cases}
\dfrac{m\bar{\phi}_c^{\,2}\gamma^{3}}{\pi r^2 (\tD + \rD)}\,u^{-5}, & u \gtrsim 1,\\[6pt]
\dfrac{m\bar{\phi}_c^{\,2}\gamma^{3}}{\pi r^2 (\tD + \rD)}\,u^{-3}
\exp\!\left(-\dfrac{2}{u^{\alpha}}\right), & u \lesssim 1,
\end{cases}
\eea
so that
\bea\label{eq: simplified form of rho g}
\rho_g(r,t) = \frac{A}{r^2 (\tD + \rD)}\, f(u)\,,
\eea
with $A = m \bar{\phi}_c^{\,2}\gamma^{3}/\pi$.

Eq.~\eqref{eq: simplified form of rho g} shows that the gradient energy density approaches a scaling solution: its amplitude decays as $r^{-2}(\tD + \rD)^{-1}$, while its shape depends only on the single dimensionless variable $u$. In the limit $\rD \gg r_c$ and $\tD \gg t_c$, the peak value scales as $\rho_{g,\text{peak}} \propto r^{-3}$, in agreement with numerical simulations of supercooled phase transitions~\cite{Lewicki:2020jiv}.

\item Shell width

We now analyze the shell width based on the gradient energy density distribution $\rho_g(r,t)$. The energy is predominantly localized near the $u=1$ surface. In this vicinity, since $\tD - \rD$ is suppressed by $\gamma^{-2}$, the variables $r$ and $(\tD + \rD)$ can be treated as approximately constant. Consequently, the profile of $\rho_g(r,t)$ depends only on $\tD - \rD$.

Eq.~\eqref{eq: simplified form of rho g} reveals that this profile is governed by a shape function $f(u)$, where $u$ is given in Eq.~\eqref{eq:u}. This dependence implies that the width of $\rho_g(r,t)$ in $\rD$ near the $u=1$ surface at time $\tD$ is set by $(\tD + \rD)/(4\gamma^2)$,
\begin{align}
    \Delta\rD \sim (\tD+\rD)/4\gamma^2\ .
\end{align}
When $\rD\gg r_c,,\tD \gg t_c$, we have $\Delta\rD$ scales as $r^1$. 

\item Shell radius

We define the shell radius $r_s$ as the location of the peak gradient energy density, which yields
\begin{align}
    r_s=r_c+\tD-\frac{\tD+\rD}{4\gamma^2}\ .
\end{align}
The expression for the bubble radius can be understood from a physical perspective. Even though the bubble wall is highly relativistic, its group velocity remains smaller than the speed of light. For a free massive scalar field with the dispersion relation $\omega^2 = m^2 + k^2$, the group velocity is $v_g = d\omega/dk = k/\omega$. The maximum momentum $k \sim \gamma m$ — and hence the largest possible $v_g$ — is determined at the moment of the initial collision. This leads directly to the result
\begin{align}
    v_g=\frac{k}{\omega}=\frac{\gamma}{\sqrt{1+\gamma^2}}\approx 1-\frac{1}{2\gamma^2}\ .
\end{align}
Then, the radius of the mode with the largest momentum $m\gamma$ reads
\begin{align}
    r_{\gamma}=r_c+\tD(1-v_g)=r_c+\tD-\frac{\tD}{2\gamma^2}\approx r_s\ ,
\end{align}
which coincides precisely with the location of the peak gradient energy density. This connection offers an alternative perspective for understanding the exponential decay of $\phi$ in the region $\rD > v \tD$: this region is causally unreachable from the modes that carry most of the shell's energy, leading to an exponential suppression.

\item{Total gradient energy per solid angle}

The total gradient energy per unit solid angle of a bubble shell is
\begin{align}
    E_g(t)=\int\rho_g(r,t) r^2 \mathrm{d}r\ .
\end{align}
From the scaling relations $\Delta r \propto r^1$ and $\rho_g \propto r^{-3}$ derived above, we conclude that $E_g$ is conserved as it should be due to energy conservation. 

Then using the expressions of $\rho_g$ at the two regimes in Eq.~\eqref{eq:rho g} and the relation between $u$ and $\rD$ and $\tD$, we can estimate $E_g$ as 
\begin{align}\label{eq:full integration of sigma 0}
    E_g(t)\approx & \frac{m\gamma\bar{\phi}_c^2}{4\pi } \left( {\cal I}_1 + {\cal I}_2 \right) \ ,
\end{align}
where ${\cal I}_1$ and ${\cal I}_2$ are the contributions from the $\rD < v \tD$ and $\rD > v \tD$ regimes, respectively. In detail, we have
\bea
{\cal I}_1 =  \int_0^{1}\frac{2\mathrm{d}u}{u^2}\exp\left[-2\left(\frac{1}{u}\right)^\alpha\right] \ , \;\;\; 
{\cal I}_2 = \int_{1}^{\infty}\frac{2\mathrm{d}u}{u^4} \ .
\eea
The integrations of both ${\cal I}_1$ and ${\cal I}_2$ are dominated in the region $u\sim 1$, indicating that the conserved quantity 
\bea
E_g = {\cal C} \frac{m\gamma\bar{\phi}_c^2}{4\pi } \ ,
\eea
with ${\cal C}$ a model dependent order one numerical constant that can be determined from numerical simulations.

\item Effects of nonlinear terms

Substituting the field configurations in Eqs.~\eqref{eq:phiL} and~\eqref{eq:phiR} into the nonlinear terms in the EoM (e.g., the $\phi^2$ term), we find that the scaling of $\phi_{\ell m}$ implies that these nonlinear contributions decay at least as $r^{-3}$. They therefore rapidly become negligible compared to the mass term as the bubble expands.

Subsequent collisions between the shells are also negligible. This is not only because the potential term is suppressed by a factor of $\gamma^{-2}$ relative to the kinetic terms, but also because the time interval during which the two shells pass through each other is further reduced by an additional factor of $\gamma$.

In this work, we quantify the impact of nonlinearities using numerical simulations in Sec.~\ref{sec:two bubble simulation}, while a complete analytical treatment is left for future study.

\end{itemize}

\subsubsection{Evolution of energy shells in radiation domination universe}

We now generalize the flat-spacetime results to a radiation-dominated universe and study the propagation of bubble shells in this background. We adopt the FRW metric given in Eq.~\eqref{eq: frw metric}. The EoM for a free massive scalar field is  
\begin{align}
    \partial_\eta^2\phi+\frac{2}{\eta}\partial_\eta\phi-\nabla^2\phi+a^2m^2\phi=0\ .
\end{align}

Following the flat-spacetime analysis, we decompose the field into spherical harmonics and define
\bea
\chi_{\ell m}=\eta r\,\phi_{\ell m}\ ,
\eea
where the additional factor of $\eta$ is introduced to absorb the Hubble friction term. Neglecting the angular derivative terms, $\chi_{\ell m}$ satisfies a $1+1$ dimensional equation,
\begin{align}
    \partial_\eta^2\chi_{\ell m}-\partial_r^2\chi_{\ell m}+a^2 m^2\chi_{\ell m}=0\ .
\end{align}

We again solve this equation using the Green’s function method. The Green’s function $G(\eta,r;\eta',r')$ obeys
\begin{align}\label{eq: greens RD}
    \left(\partial_\eta^2-\partial_r^2+a^2m^2\right)G(\eta,r;\eta',r')
    =\delta(\eta-\eta')\delta(r-r')\ .
\end{align}
Applying a Fourier transform with respect to $r$, we define
\begin{align}
    \tilde{g}(k,\eta,\eta')=\int \mathrm{d}r\, e^{-ik(r-r')}G(\eta,r;\eta',r')\ .
\end{align}
Equation~\eqref{eq: greens RD} then becomes
\begin{align}
    \left(\partial_\eta^2+k^2+a^2m^2\right)\tilde{g}(k,\eta,\eta')
    =\delta(\eta-\eta')\ .
\end{align}

Due to the explicit time dependence of the mass term, the Green’s function cannot be obtained analytically. We therefore employ the WKB approximation to solve the mode equation. The leading-order WKB solutions are
\begin{align}
    u_{\pm}(\eta)=\frac{1}{\sqrt{2\omega_k(\eta)}}
    \exp\!\left(\pm i\int_{\eta_0}^{\eta}\omega_k(\eta'')\,\mathrm{d}\eta''\right)\ ,
\end{align}
where
\bea
\omega_k(\eta)=\sqrt{k^2+m^2a^2(\eta)}\ .
\eea
The corresponding retarded Green’s function is then
\begin{align}
    \tilde{g}(k,\eta,\eta')
    =\frac{1}{\sqrt{\omega_k(\eta)\omega_k(\eta')}}
    \sin\!\left(\int_{\eta'}^{\eta}\omega_k(\eta'')\,\mathrm{d}\eta''\right)
    \theta(\eta-\eta')\ .
\end{align}

Performing the inverse Fourier transform yields
\begin{align}
    G(\eta,r;\eta',r')
    =\theta(\eta-\eta')
    \int\frac{\mathrm{d}k}{2\pi}e^{ik(r-r')}
    \frac{1}{\sqrt{\omega_k(\eta)\omega_k(\eta')}}
    \sin\!\left(\int_{\eta'}^{\eta}\omega_k(\eta'')\,\mathrm{d}\eta''\right)\ .
\end{align}
This integral can be evaluated using the saddle-point approximation. Writing the sine function in exponential form, we obtain
\begin{align}
    G(\eta,r;\eta',r')
    =\theta(\eta-\eta')\frac{I_+-I_-}{2i}\ ,
\end{align}
with
\begin{align}
    I_{\pm}
    =\int\frac{\mathrm{d}k}{2\pi}
    \frac{1}{\sqrt{\omega_k(\eta)\omega_k(\eta')}}
    \exp\!\left[i k(r-r')\pm i\int_{\eta'}^{\eta}\omega_k(\eta'')\,\mathrm{d}\eta''\right]\ .
\end{align}

Defining the phase
\bea
\Phi_{\pm}(k)=k(r-r')\pm \int_{\eta'}^{\eta}\omega_k(\eta'')\,\mathrm{d}\eta''\ ,
\eea
the saddle points $k^*_{\pm}$ satisfy
\begin{align}\label{eq: line of sight condition}
    \frac{\mathrm{d}\Phi_{\pm}}{\mathrm{d}k}=0
    \quad\Rightarrow\quad
    (r-r')\pm \int_{\eta'}^{\eta}\frac{\mathrm{d}\omega_k(\eta'')}{\mathrm{d}k}\,\mathrm{d}\eta''=0\ .
\end{align}
Expanding the integrand around the saddle point and carrying out the Gaussian integral, we find
\begin{align}
    I_{\pm}\approx
    \frac{1}{\sqrt{2\pi}}
    \frac{1}{\sqrt{\omega_{k^*_{\pm}}(\eta)\omega_{k^*_{\pm}}(\eta')}}
    \frac{1}{\sqrt{|\Phi_{\pm}''(k^*_{\pm})|}}
    \exp\!\left[i\Phi_{\pm}(k^*_{\pm})
    +i\frac{\pi}{4}\,\mathrm{sgn}\,\Phi_{\pm}''(k^*_{\pm})\right]\ ,
\end{align}
where
\begin{align}
    \Phi_{\pm}''(k^*_{\pm})
    =\left.\pm \int_{\eta'}^{\eta}
    \frac{\mathrm{d}^2\omega_k(\eta'')}{\mathrm{d}k^2}\,
    \mathrm{d}\eta''\right|_{k=k^*_{\pm}}\ .
\end{align}

For $r-r'>0$, and noting that $\omega_k$ is an even function of $k$, the saddle points satisfy $k_+^*=-k_-^*<0$. Defining $k^*\equiv k_-^*>0$, the Green’s function becomes
\begin{align}\label{eq: frw greens function method}
    G(\eta,r;\eta',r')
    =\theta(\eta-\eta')
    \frac{1}{\sqrt{2\pi\omega_{k^*}(\eta)\omega_{k^*}(\eta')\Phi_{+}''(k^*)}}
    \cos\!\left[-k^*(r-r')
    +\int_{\eta'}^{\eta}\omega_{k^*}(\eta'')\,\mathrm{d}\eta''
    -\frac{\pi}{4}\right]\ .
\end{align}

As a consistency check, we consider the case $a=1$, which reduces to flat spacetime. In this limit,
\begin{align}
    k^*=m\,\frac{r-r'}{\sqrt{(\eta-\eta')^2-(r-r')^2}}\ ,
\end{align}
and the Green’s function reduces to
\begin{align}\label{eq:flat green expansion}
    G(\eta,r;\eta',r')
    =\theta(\eta-\eta')
    \frac{1}{\sqrt{2\pi m s}}
    \cos\!\left(ms-\frac{\pi}{4}\right)\ ,
\end{align}
which is precisely the asymptotic expansion of the flat-spacetime Green’s function. This demonstrates that the saddle-point approximation is valid provided $ms$ is not too small.

The Green’s function in its full form is rather involved: for given $r$ and $r'$, one must first determine the saddle point $k^*$ from Eq.~\eqref{eq: line of sight condition} and then substitute it into Eq.~\eqref{eq: frw greens function method}. However, the modes relevant for our analysis have comoving momentum $k\sim\gamma m\gg a m$. In this high-momentum regime, Eq.~\eqref{eq: line of sight condition} admits a simple analytical approximation,
\begin{align}
(\eta-\eta')-(r-r')
=\frac{m^2}{2k^{*2}}\int_{\eta'}^{\eta}a^2(\eta'')\,\mathrm{d}\eta''
=\frac{m^2\,\mathcal{R}(\eta,\eta')}{4k^{*2}}\ .
\end{align}
Here we have introduced the function
\begin{align}
\mathcal{R}(\eta,\eta')
\equiv 2\int_{\eta'}^{\eta}a^2(\eta'')\,\mathrm{d}\eta''\ .
\end{align}

Using this relation, the Green’s function in Eq.~\eqref{eq: frw greens function method} simplifies to
\begin{align}
G(\eta,r;\eta',r')
=\theta(\eta-\eta')
\frac{
\cos\!\left(
m\sqrt{\mathcal{R}(\eta,\eta')\big[(\eta-\eta')-(r-r')\big]}
-\frac{\pi}{4}
\right)
}{
\sqrt{
2\pi m\sqrt{\mathcal{R}(\eta,\eta')\big[(\eta-\eta')-(r-r')\big]}
}
}\ .
\end{align}

Comparing with the flat-spacetime Green’s function in Eq.~\eqref{eq:flat green expansion}, we see that the radiation-dominated result is obtained by the simple replacement of the flat-spacetime invariant interval with the effective quantity $\mathcal{R}(\eta,\eta')$, encoding the cumulative effect of cosmic expansion.

We again define the shifted coordinates $\rD = r - r_c$ and $\tilde\eta = \eta - \eta_c$. The corresponding generalization of Eq.~\eqref{eq:third form} is  
\begin{align}
    \chi_{\ell m}(r,\eta)
    = -\sqrt{\frac{2}{\pi}}
    \int \mathrm{d}\rp\,
    \frac{
        \cos\!\left(
            m\sqrt{\mathcal{R}(\tilde\eta)\big(\tilde\eta - \rD + \rp\big)}
            - \frac{\pi}{4}
        \right)
    }{
        \sqrt{
            m\sqrt{\mathcal{R}(\tilde\eta)\big(\tilde\eta - \rD + \rp\big)}
        }
    }
    \,\partial_{\rp}\phi_c(\rp)\,.
\end{align}
Here $\mathcal{R}(\tilde\eta)\equiv\mathcal{R}(\eta_c,\eta)$. Analogous to the flat-spacetime case, the gradient energy density can be computed and is given by  
\begin{align}\label{eq:rho g 1}
    \rho_g(r,\eta) &=
    \begin{cases}
        \dfrac{m\bar{\phi}_c^2\gamma^3}{\pi r^2\eta^2\mathcal{R}(\tilde\eta)}\,u^{-5}\,, & u\gtrsim 1\,, \\[8pt]
        \dfrac{m\bar{\phi}_c^2\gamma^3}{\pi r^2\eta^2\mathcal{R}(\tilde\eta)}\,
        u^{-3}\exp\!\left(-\dfrac{2}{u^{\alpha}}\right)\,, & u\lesssim 1\,,
    \end{cases}
\end{align}
with
\bea
    u = 2\gamma \frac{\sqrt{\tilde\eta - \rD}}{\sqrt{\mathcal{R}(\tilde\eta)}}\,.
\eea

Compared with the flat-spacetime result, we observe two key modifications in a radiation-dominated universe: (i) the time–distance combination appearing in the flat solution is effectively replaced by $\mathcal{R}(\tilde\eta)$; (ii) an extra factor of $\eta^{-2}$ appears in the denominator due to Hubble friction. As in the flat case, the gradient energy density of the bubble shell peaks along the worldline of the cutoff mode $k = \gamma m$, and decays as  
\begin{align}\label{eq:gradient after collision}
    \rho_g(r,\eta)\propto \frac{1}{r^2\eta^2 \mathcal{R}(\tilde\eta)}\,.
\end{align}
The comoving width of the bubble shell scales as $\Delta r \sim \mathcal{R}(\tilde\eta)$, and the total gradient energy per solid angle scales as  
\begin{align}\label{eq:tension after collision}
    E_g(\eta)\propto \eta^{-2}\,.
\end{align}

In this case, the field amplitude scales as $\phi \sim r^{-1}\eta^{-1}\mathcal{R}^{-1/2}(\tilde\eta)$, which decays faster than in flat spacetime. Consequently, nonlinear terms are expected to decouple earlier during radiation domination. We can also determine when the thin-shell approximation breaks down in an expanding universe. In the late-time limit $\eta \gg \eta_c$, the ratio of the shell width to its radius is  
\begin{align}
    \frac{\Delta r}{r_{\mathrm{bub}}}
    \sim \frac{\mathcal{R}(\eta)}{\gamma^2 \eta}
    \sim \left(\frac{\eta}{\eta_c\gamma}\right)^2\,.
\end{align}
Therefore, the thin-shell approximation breaks down when $a(\eta)/a(\eta_c) \sim \gamma$, which is equivalent to the cutoff momentum of the bubble wall becoming non-relativistic. These scaling properties will be useful in our discussion of the long-term gravitational-wave source.

\subsection{False vacuum fraction and background evolution}
\label{sec:fv fraction}

In this subsection we discuss the evolution of the false-vacuum fraction and the resulting constraint on the energy released during the phase transition.

Since the wall velocity before collision is $v_w \simeq 1$, the comoving radius of a bubble nucleated at time $t'$ and observed at time $t$ is
\bea
R(t,t') = \int_{t'}^t a^{-1}(t'')\,\mathrm{d}t'' \,.
\eea
The probability for a point in space to still reside in the false vacuum $\phi_F$ at time $t$ is then
\begin{align}
{\cal P}_{\rm false}(t)
= \exp\left[
- \frac{\Gamma}{\mathcal{V}}
\int_0^t \mathrm{d}t'\,
\frac{4\pi}{3} R(t,t')^3 a^3(t')
\right]\!,
\label{eq:false vacuum retio}
\end{align}
which equals the fraction of the volume that remains in the false vacuum.

During radiation domination, $a(t'') = a(t')\,(t''/t')^{1/2}$. Hence
\bea
R(t,t') = 2\,a^{-1}(t')\big(\sqrt{t t'} - t'\big)\,,
\eea
and one finds
\bea
{\cal P}_{\rm false}(t)
= \exp\left(
- \frac{\Gamma}{\mathcal{V}}\,
\frac{8\pi}{105}\,t^4
\right)\!.
\eea
We define the percolation time $t_p$ by ${\cal P}_{\rm false}(t_p) = 0.71$; gravitational waves produced before $t_p$ are negligible \cite{Athron:2023mer,LIN2018299}. This gives
\bea
t_p \simeq 1.1\left(\frac{\Gamma}{\mathcal{V}}\right)^{-1/4},
\qquad
H_p \equiv \frac{1}{2t_p}
= 0.46\left(\frac{\Gamma}{\mathcal{V}}\right)^{1/4}.
\eea

The false-vacuum energy density remains constant, while the radiation density redshifts as $a^{-4}$. To include the effect of vacuum energy on the expansion history, we split the total energy density into radiation and vacuum components and write the Friedmann equation as
\bea
3H^2 M_{\rm Pl}^2 = \rho_r + \rho_v\,.
\eea
Vacuum energy is converted into radiation via bubble nucleation and expansion, leading to
\begin{align}\label{eq: background eom}
\dot{\rho}_r + 4H\rho_r = -\dot{\rho}_v\,.
\end{align}

In our setup, the transition is characterized by the nucleation rate $\Gamma/\mathcal{V}$ and the latent heat $\Delta\rho$. We introduce the dimensionless parameter
\bea
\xi
= \sqrt{\frac{8\pi G_N}{3}\,\Delta\rho}\,
\left(\frac{\Gamma}{\mathcal{V}}\right)^{-1/4}.
\eea
Percolation requires that the transition completes before vacuum domination. Using $H_p = 0.46(\Gamma/\mathcal{V})^{1/4}$, if $\xi > 0.46$ the universe enters an eternal inflation regime where true-vacuum regions remain disconnected until vacuum domination. To maintain radiation domination during the transition we therefore require at least $\xi \lesssim 0.5$.

A stronger bound follows from the constraint on $\Delta N_{\rm eff}$. Today,
\bea
\left.\frac{\rho_{\rm DS}}{\rho_\gamma}\right|_{\rm today}
= \frac{g_{*{\rm DS}}}{2}
\left(\frac{T_{\rm DS}}{T_{\rm CMB}}\right)^4
= \frac{7}{8}\left(\frac{4}{11}\right)^{4/3}\Delta N_{\rm eff}\,,
\eea
where $\rho_{\rm DS}$ is the dark-sector energy density. Assuming the number of relativistic degrees of freedom in the dark sector does not change, at the end of the phase transition we obtain
\bea
\left.\frac{\rho_{\rm DS}}{\rho_{\rm SM}}\right|_{\rm end}
= 2\left(\frac{11}{43}\right)^{4/3} g_*^{1/3}
\left.\frac{\rho_{\rm DS}}{\rho_\gamma}\right|_{\rm today}
= \frac{7}{4}\left(\frac{4}{43}\right)^{4/3}
g_*^{1/3}\Delta N_{\rm eff}\,.
\eea
Taking $g_* = 106.75$ and $\Delta N_{\rm eff} < 0.4$ \cite{DESI:2024mwx,Allali:2024cji} implies
\bea
\left.\frac{\rho_{\rm DS}}{\rho_{\rm SM}}\right|_{\rm end} \lesssim 0.14\,.
\eea
Solving Eq.~\eqref{eq: background eom} one finds that the dark-sector energy fraction is related to $\xi$ approximately by
\bea
\frac{\rho_{\rm DS}}{\rho_{\rm SM}} \simeq 7.3\,\xi^2\,,
\eea
so the above bound yields
\bea
\xi \lesssim 0.14\,.
\eea

\begin{figure}[htpb]
\centering
\includegraphics[width=0.6\linewidth]{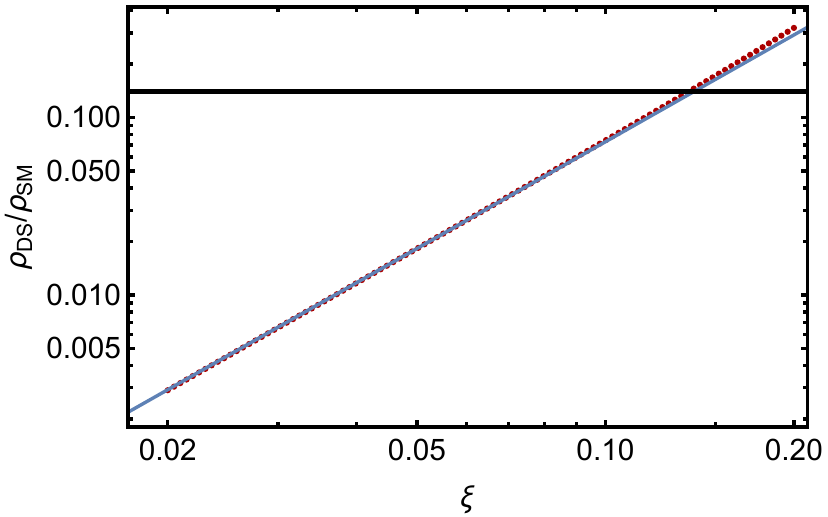}
\caption{The relation between $\xi$ and the energy fraction of DS. The red dotted line is the result from simulation while the blue solid line is the fitting result $\frac{\rho_{\rm DS}}{\rho_{\rm SM}}\simeq 7.3\xi^2$. The $\Delta N_{\rm eff}$ bound is shown in the black horizontal line.}
\label{fig: dne}
\end{figure}

\section{Qualitative Features of the GWs produced by Dark Meta Stable Vacuum Decay}
\label{sec:GW}

Including the metric perturbations in the spatial part, the line element can be written as 
\bea
\D s^2 = a^2 (\D\eta^2 - (\delta_{ij} + \hij) \D x^i \D x^j) \ ,
\eea
and the GWs are identified with the transverse-traceless part $h^{\rm TT}_{ij}$, which obeys the linearized Einstein equation
\bea\label{eq:eom1}
h^{\rm TT''}_{ij} + 2{\cal H} h^{\rm TT'}_{ij} - \nabla^2 h^{\rm TT}_{ij} = 16 \pi G_N T_{ij}^{\rm TT} \ ,
\eea
where ${\rm TT}$ denotes the transverse and traceless components, ${\cal H} = a'/a$, $T_{ij}$ is the spatial part of the energy-momentum tensor. Here a prime denotes the derivative with respect to conformal time $\eta$. For convenience, define the rescaled tensor mode 
\bea
\h_{ij}(\eta,\bx) = a(\eta) h_{ij}^{\rm TT}(\eta,\bx) \ .
\eea
Then Eq.~(\ref{eq:eom1}) becomes 
\bea\label{eq:eom2}
\h''_{ij} - \left( \nabla^2 + \frac{a''}{a} \right) \h_{ij} = 16 \pi \gn a(\eta) T^{\rm TT}_{ij}(\eta,\bx) \ .
\eea
During radiation domination, $a$ is linear in $\eta$ so that $a''=0$. The Fourier modes of $\h$ satisfy
\bea\label{eq:GW eom}
\trh''_{ij}(\eta,\bk) + k^2 \trh_{ij}(\eta,\bk) = 16 \pi G_N a(\eta) \tTT(\eta,\bk) \ .
\eea
Using the Green's function method, we obtain
\bea\label{eq:greens function}
\trh_{ij}(\eta,\bk) = 16 \pi \gn \int d\eta' \frac{1}{k} \sin k(\eta-\eta') a(\eta') \tTT(\eta',\bk) \ .
\eea
Deeply inside the horizon, the GW energy density can be written as
\bea\label{eq:rho0}
\rho_{\rm GW} &=& \frac{1}{32\pi \gn a^4(\eta)V}\sum_{ij} \int \frac{d^3 k}{(2\pi)^3}|\partial_\eta\trh_{ij} |^2\nn
&=& \frac{4\pi G_N}{a^4(\eta) V} \sum_{ij} \int \frac{d^3 k}{(2\pi)^3} \left[ \left|{I_c}_{ij}(\bk)\right|^2 + \left|{I_s}_{ij}(\bk)\right|^2 \right] \ ,
\eea
where 
\bea\label{eq:IcIs}
{I_c}_{ij}(\bk) &=&  \int_{\eta_i}^{\eta_f} d\eta'\cos(k\eta')a(\eta')\tTT(\eta',\bk) \ , \\
{I_s}_{ij}(\bk) &=&  \int_{\eta_i}^{\eta_f} d\eta'\sin(k\eta')a(\eta')\tTT(\eta',\bk) \ .
\eea
Here $\eta_i$ and $\eta_f$ denote the times the GW production starts and ends, respectively. The quantity $V$ is the total comoving volume; it arises from the Fourier-mode normalization and cancels in the final result. 
$\rho_{\rm GW}$ can further be written as
\bea\label{eq:rho}
\rho_{\rm GW} &=& \frac{4\pi \gn}{a^4(\eta) V} \sum_{ij} \int \frac{d^3 k}{(2\pi)^3} \int_{\eta_i}^{\eta_f} \D \eta' \D \eta'' a(\eta') a(\eta'') \cos[k(\eta'-\eta'')] \tTT(\eta',\bk) \tilde{T}^{\rm TT*}_{ij} (\eta'',\bk)  \ .\nn 
\eea

As shown in Sec.~\ref{sec:meta}, the bubble wall thickness and is significantly smaller than the typical bubble radius, and right after the collision, the Lorentz boost factor $\gamma\gg1$, and these properties last for several e-folds. In this section, we analyze the qualitative features of the GW spectrum such as the slope of the infrared part of the spectrum and the strength and frequency of the GW signal today using the thin-wall and large-boost approximation. We leave the full numerical result to the next section.

\subsection{The thin wall and large boost approximation}

In metastable dark sector vacuum decay, most of the energy released during the phase transition is carried by the bubble walls before collision or bubble shells after collision, as discussed in Sec.~\ref{sec:meta}. To calculate the bubble-wall contribution to the total $\tTT$, we decompose $\tilde{T}^{\rm TT}_{ij}$ into a sum over individual bubbles
\begin{align}\label{eq: tij decompose into bubbles}
    \tilde{T}^{\rm TT}_{ij} (\eta,\bk)=\sum_{n}e^{-i \bk\cdot \mathbf{r}_n }\Lambda_{ij,kl}(\hat{\bk})\int \mathrm{d}x^3e^{-i\bk\cdot \mathbf{x}}T^{B}_{ij,n}(\eta,\mathbf{x}) \ ,
\end{align}
where $\mathbf{r}_n$ denotes the center of the $n$th bubble and $T^{B}_{ij,n}(\eta,\mathbf{x})$ is the energy-momentum tensor of the $n$th bubble. In writing Eq.~\eqref{eq: tij decompose into bubbles}, we have shifted the integration origin to each bubble's center.

As discussed in Sec.~\ref{sec:meta}, the bubble walls are much thinner than the relevant dark sector scales, so we use the thin wall approximation for $T^B_{ij,n}$.
Introducing spherical coordinates $(r,\theta,\psi)$ centered on a bubble center, the field gradient of the bubble configuration $\phi_B$ reads 
\bea
\nabla \phi_B=\partial_r\phi_B\hat{e}_r+\frac{\partial_\theta\phi_B}{r} \hat{e}_\theta+\frac{\partial_\psi\phi_B}{r\sin\theta}\hat{e}_\psi \ .
\eea
Because $\partial_r\phi_B\sim \gamma m\phi_B$ dominates the angular derivatives, we may approximate $\partial_{i}\phi_B\approx\hat{x}_i\partial_r\phi_B$, where $\gamma$ is the Lorentz boost factor of the bubble wall. 

Each point on the $n$th bubble wall eventually collides with other bubbles at some time $\eta_c$. We denote the collision-time distribution over the wall of the $n$th bubble by $\eta_c^{(n)}(\theta,\psi)$. For a point labeled by $(\theta,\psi)$, prior to collision it follows the uncollided wall evolution given by Eq.~\eqref{eq: bubble eff eom}. Because bubble walls propagate at nearly the speed of light, a given patch remains causally unaware of whether other regions of the bubble have broken. Consequently, until it collides, the local energy–momentum tensor retains the uncollided-wall form, even if other patches have already collided:
\begin{align}
    T^{B}_{ij,n}(\eta,r,\theta,\psi)=\hat{x}_i\hat{x}_j a^2\gamma^2(\eta)\phi_s'^2\ .
\end{align}
Here $\phi_s$ is the wall profile in the rest frame defined in Eq.~\eqref{eq:bubble wall ansatz}. After collision, as shown in Sec.~\ref{sec:meta}, due to the large boost $\gamma$, the field configuration soon becomes thin-wall like. Thus, we can still use the thin wall approximation to estimate the energy-momentum tensor. This treatment is referred as the bulk flow model~\cite{Konstandin:2017sat,Jinno:2017fby}. While Refs.~\cite{Konstandin:2017sat,Jinno:2017fby} discuss flat spacetime, here the expansion of the Universe is non-negligible; therefore, additional expansion-induced decay factors appear. Using Eq.~\eqref{eq:gradient after collision}, we obtain
\begin{align}
    T^{B}_{ij,n}(\eta,r,\theta,\psi)\propto\hat{x}_i\hat{x}_j \eta^{-2}r^{-2}\mathcal{R}^{-1}[\eta-\eta_c^{(n)}(\theta,\psi)],\qquad \eta>\eta_c^{(n)}(\theta,\psi) \ .
\end{align}

Under the thin-wall and large-momentum approximation, we can carry out the radial integral appearing in the Fourier transform in Eq.~\eqref{eq: tij decompose into bubbles}. Before collision,
\begin{align}\label{eq: before c tij inter}
    \int r^2\mathrm{d}rT^{B}_{ij,n}(\eta,\mathbf{x})&=\hat{x}_i\hat{x}_j \left(\eta-\eta^{(n)}_0\right)^2 \int \mathrm{d}ra^2\gamma^2(\eta)\phi_s'^2\\
    &=\hat{x}_i\hat{x}_j \left(\eta-\eta^{(n)}_0\right)^2  a\gamma(\eta)\sigma\\
    &=\hat{x}_i\hat{x}_j \left(\eta-\eta^{(n)}_0\right)^3  a^2 \delta V \ ,
\end{align}
where $\eta^{(n)}_0$ is the nucleation time of the $n$th bubble, $\sigma$ is the wall tension, and $\delta V$ is the latent energy density. After collision, using Eq.~\eqref{eq:tension after collision}
\begin{align}\label{eq: after c tij inter}
    \int r^2\mathrm{d}rT^{B}_{ij,n}(\eta,\mathbf{x}) &\propto \hat{x}_i\hat{x}_j  \eta^{-2}
\end{align}

After collision, $\mathcal{O}(1)$ fraction of energy is carried by bubble walls. Here we simply assume the continuity of the wall energy-momentum tensor across the collision, i.e., equating the expressions above at $\eta=\eta_c^{(n)}$. Eq.~\eqref{eq: after c tij inter} then yields
\begin{align}
    \int r^2\mathrm{d}rT^{B}_{ij,n}(\eta,\mathbf{x})=\hat{x}_i\hat{x}_j(\eta_c^{(n)}-\eta^{(n)}_0)^3  a^2 (\eta_c^{(n)})\delta V \left(\frac{\eta_c^{(n)}}{\eta}\right)^2 \ .
\end{align}
Substituting these results into Eq.~\eqref{eq: tij decompose into bubbles} and choosing the $z$ axis parallel to $\mathbf{k}$, we find
\begin{align}\label{eq:signle bubble tij before collision}
\tilde{T}^{\rm TT}_{ij} (\eta,\bk)=\delta V\sum_{n}e^{-i \bk\cdot \mathbf{r}_n }\Lambda_{ij,kl}(\hat{\bk})\int \mathrm{d}\Omega \hat{x}_i\hat{x}_j e^{-ik\left(\eta-\eta^{(n)}_0\right)\cos\theta}f(\eta,\eta^{(n)}_0)
\end{align}
with
\begin{align}
    f(\eta,\eta^{(n)}_0)= \left(\eta-\eta^{(n)}_0\right)^3  a^2(\eta) \ ,
\end{align}
for $\eta<\eta_c^{(n)}(\theta,\psi)$, and, for $\eta\geq\eta_c^{(n)}(\theta,\psi)$, 
\bea
f(\eta,\eta^{(n)}_0) = (\eta_c^{(n)}-\eta^{(n)}_0)^3  a ^4(\eta_c^{(n)})a^{-2}(\eta) \ .
\eea

For comparison, if we consider the envelop model, which assumes the patch of bubble wall vanishes right after collision, we will have $f(\eta,\eta^{(n)}_0) = 0$ for $\eta \geq \eta_c^{(n)}(\theta,\psi)$.

\subsection{The infrared part of the GW signal}

The IR GW spectrum is typically insensitive to the detailed particle-physics model of the phase-transition sector. For instance, if the source has no super-horizon spatial correlations (correlation length $\lesssim H_{\mathrm{PT}}^{-1}$) and the bubble configuration persists for less than a Hubble time, the IR slope scales as $k^3$~\cite{Caprini:2009fx,Cai:2019cdl}. To see this, assume that $k$ is much smaller than all characteristic energy scales of the system so that $\tilde{T}_{ij}(\eta,\mathbf{k}) \simeq \tilde{T}_{ij}(\eta,\mathbf{0})$. Under this approximation, Eq.~\eqref{eq:rho} gives
\begin{align}
\frac{\mathrm{d}\rho_{\rm GW}^{\rm IR}}{\mathrm{d}\ln k}
= \frac{2 G_N k^3}{\pi a^4(\eta)\,{\cal V}}
\sum_{ij} \int_{\eta_i}^{\eta_f} \mathrm{d}\eta'\,\mathrm{d}\eta''\,
a(\eta') a(\eta'')\,
\big\langle \tilde{T}_{ij}(\eta',\mathbf{0})\,
\tilde{T}^*_{ij}(\eta'',\mathbf{0})\big\rangle \,.
\end{align}
The time integral is $k$-independent, so the $k^3$ scaling arises purely from the phase-space factor. Consequently, the IR power spectrum grows as $k^3$.

In the case of dark-sector metastable vacuum decay, however, the situation is different. Since $\gamma \gg 1$, the kinetic and gradient energies of the wall dominate over the potential energy, and the bubble-wall evolution approximately follows the free-field EoM. The walls therefore persist for much longer than a Hubble time, and the naive $k^3$ IR slope can be modified.

To examine the IR behavior more carefully, we take a closer look at the integral $I_c$ defined in Eq.~\eqref{eq:IcIs},
\begin{align}\label{eq: integration for tij}
I_c(\mathbf{k})
= \int_{\eta_i}^{\eta_f} \mathrm{d}\eta'\,
\cos(k\eta')\,a(\eta')\,\tilde{T}_{ij}(\eta',\mathbf{k}) \,,
\end{align}
where we have suppressed the explicit spatial indices $i,j$ for brevity.

To qualitatively evaluate $I_c$, we introduce $\eta_b$, the characteristic time by which essentially all collisions have completed, $\eta_b \sim H^{-1}$. We then split the integral into early and late contributions,
\begin{align}
I_c
&= I_{c,0} + I_{c,\ell} \nonumber\\
&= \int_{\eta_i}^{\eta_b} \mathrm{d}\eta'\,
\cos(k\eta')\,a(\eta')\,\tilde{T}_{ij}(\eta',\mathbf{k})
+ \int_{\eta_b}^{\eta_f} \mathrm{d}\eta'\,
\cos(k\eta')\,a(\eta')\,\tilde{T}_{ij}(\eta',\mathbf{k}) \,.
\end{align}
Because of the large Lorentz boost factor $\gamma$, $\tilde{T}_{ij}$ does not change significantly within one e-fold. We therefore expect $I_{c,0}$ to be subdominant compared to $I_{c,\ell}$. In what follows we focus on modes with
\[
\eta_f^{-1} \ll k \ll H_{\mathrm{PT}}\,,
\]
since for $k \ll \eta_f^{-1}$ the above argument already implies a $k^3$ spectrum.

Applying the thin-wall approximation with the single-bubble configuration in Eq.~\eqref{eq:signle bubble tij before collision}, the late-time contribution can be written as
\begin{align}\label{eq: Icl}
I_{c,\ell}
= \delta V \sum_n e^{-i\mathbf{k}\cdot\mathbf{r}_n}\,
\Lambda_{ij,kl}(\hat{\mathbf{k}})
\int_{\eta_b}^{\eta_f} \mathrm{d}\eta'\,
\cos(k\eta')\,a(\eta')
\int \mathrm{d}\Omega\,\hat{x}_k \hat{x}_l\,
e^{-ik(\eta' - \eta_0^{(n)})\cos\theta}\,
f(\eta',\eta_0^{(n)}) \,,
\end{align}
where $\mathbf{r}_n$ and $\eta_0^{(n)}$ are the nucleation position and time of the $n$th bubble, and $\delta V$ is the comoving volume element.

Since $r_n, \eta_0^{(n)} \sim H_{\mathrm{PT}}^{-1}$, we have
$e^{i\mathbf{k}\cdot\mathbf{r}_n} \simeq 1$ and
$e^{ik\eta_0^{(n)}\cos\theta} \simeq 1$ for modes with $k \ll H_{\mathrm{PT}}$. After $\eta_b$, all bubbles have been broken, so
\[
f(\eta',\eta_0^{(n)})
= \big(\eta_c^{(n)} - \eta_0^{(n)}\big)^3
a^4(\eta_c^{(n)})\,a^{-2}(\eta')\,.
\]
Consequently, Eq.~\eqref{eq: Icl} simplifies to
\begin{align}
I_{c,\ell}
= \delta V \sum_n \Lambda_{ij,kl}(\hat{\mathbf{k}})
\int_{\eta_b}^{\eta_f} \mathrm{d}\eta'\,
\frac{\cos(k\eta')}{a(\eta')}
\int \mathrm{d}\Omega\,\hat{x}_k \hat{x}_l\,
e^{-ik\eta'\cos\theta}\,
\big(\eta_c^{(n)} - \eta_0^{(n)}\big)^3
a^4(\eta_c^{(n)}) \,.
\end{align}
For the modes of interest with $k \gg \eta_f^{-1}$, we may approximate the upper limit of the $\eta'$-integral by $\infty$. Using
\begin{align}
\int_{\eta_b}^{\infty} \mathrm{d}\eta'\,
\frac{\cos(k\eta')}{a(\eta')}\,
e^{-ik\eta'\cos\theta}
= -\frac{1}{2H_{\mathrm{PT}}}
\left[
\mathrm{Ei}\big(-i(\cos\theta-1)k\eta_b\big)
+ \mathrm{Ei}\big(-i(\cos\theta+1)k\eta_b\big)
\right] ,
\end{align}
where $\mathrm{Ei}(z) = -\int_{-z}^{\infty} \mathrm{d}t\,e^{-t}/t$ is the exponential integral, we obtain
\begin{align}\label{eq: Icl logk}
I_{c,\ell}
&\approx -\frac{\delta V}{H_{\mathrm{PT}}}
\sum_n \Lambda_{ij,kl}(\hat{\mathbf{k}})
\int \mathrm{d}\Omega\,\hat{x}_k \hat{x}_l\,
\big(\eta_c^{(n)} - \eta_0^{(n)}\big)^3
a^4(\eta_c^{(n)}) \nonumber\\
&\quad\times
\left[
\log(k\eta_b)
+ \log|\sin\theta|
+ \gamma_E
+ \mathcal{O}(k\eta_b)
\right] ,
\end{align}
where we have expanded $\mathrm{Ei}(x)$ for $k\eta_b \ll 1$ and $\gamma_E$ is the Euler–Mascheroni constant.

From Eq.~\eqref{eq: Icl logk}, one sees that in the range
$\eta_f^{-1} \ll k \ll H_{\mathrm{PT}}$ the logarithmic term dominates. Noting that
\begin{align}
k\eta_b
= \frac{k a_b}{H_{\mathrm{PT}}}
= \frac{k a_b^{-1}}{H_{\mathrm{PT}}/a_b^2}
= \frac{k_{\rm phy}}{H_b}
\simeq \frac{k_{\rm phy}}{H_{\mathrm{PT}}}\,,
\end{align}
we conclude that, in the momentum interval
$\eta_f^{-1} \ll k \ll H_{\mathrm{PT}}$, the GW spectrum scales as
\begin{align}
\frac{\mathrm{d}\rho_{\rm GW}^{\rm IR}}{\mathrm{d}\ln k}
\propto k^3 \log^2\!\left(\frac{k}{H_{\mathrm{PT}}}\right) .
\end{align}
This logarithmic dependence softens the pure $k^3$ IR scaling and is a distinctive feature of GWs sourced by metastable vacuum decay. Notice that this IR scaling is also different from the $k^1$ scaling predicted by the bulk flow model in flat spacetime~\cite{Konstandin:2017sat,Jinno:2017fby}. This is because, in the radiation-dominated era, Hubble friction drives the wall energy to decay even when the bubble wall propagates freely. In contrast, for the bulk flow model in flat spacetime, the total energy of the bubble wall is conserved after collision.

\subsection{Estimation of the frequency and strength of the GW signal}

From Eqs.~\eqref{eq:rho0}, \eqref{eq:IcIs}, and~\eqref{eq: Icl logk}, we see that the collision-time function $\eta_c^{(n)}(\theta,\psi)$ plays a central role in determining the GW amplitude. It determines the energy carried by the bubble wall at the moment of collision and therefore controls the overall strength of the GW source. In particular, in Eq.~\eqref{eq: Icl logk} the amplitude of $I_{c,\ell}$ depends on the factor $\delta V\,(\eta_c^{(n)} - \eta_0^{(n)})^3 a^3(\eta_c^{(n)})$, which encodes the energy absorbed by the bubble wall from the vacuum.

For a bubble nucleated at conformal time $\eta_{0}$ and position $\mathbf{r}_0$, different patches of the wall collide at different times. Their collision-time distribution $\eta_c(\theta,\phi)$ is characterized by a probability distribution function, which we denote by $P(\eta_c,\eta_0)$. By isotropy, this distribution is uniform over solid angle.

To compute $P(\eta_c,\eta_0)$, we first define $\tilde{P}(\eta,\eta_0)$ as the probability that a wall patch nucleated at $(\eta_0,\mathbf{r}_0)$ propagates to $(\eta,\mathbf{r})$, with $|\mathbf{r}-\mathbf{r}_0| = \eta - \eta_0$, without colliding with any other bubble. Equivalently, no other bubbles have nucleated inside the past light cone of $(\eta,\mathbf{r})$; otherwise, the patch would collide before reaching this point. This probability is
\begin{align}
    \tilde{P}(\eta,\eta_0)
    = \exp\left\{
    -\frac{4\pi\Gamma}{3\mathcal{V}}
    \left[
        \int_0^{\eta}\mathrm{d}\eta'\,a^4(\eta')(\eta-\eta')^3
        - \int_0^{\eta_0}\mathrm{d}\eta'\,a^4(\eta')(\eta_0-\eta')^3
    \right]
    \right\}.
\end{align}
The second integral subtracts the contribution from the past light cone of $(\eta_0,\mathbf{r}_0)$; this subtraction enforces the condition that a bubble is already known to have nucleated at $(\eta_0,\mathbf{r}_0)$.

The quantity $\tilde{P}(\eta,\eta_0)$ also represents the fraction of bubble-wall patches that have not yet been hit by other bubbles at time $\eta$. The collision-time probability distribution is then obtained by differentiating $\tilde{P}$:
\begin{align}
    P(\eta_c,\eta_0)
    \equiv -\left.\frac{\mathrm{d}}{\mathrm{d}\eta}\tilde{P}(\eta,\eta_0)\right|_{\eta=\eta_c} .
\end{align}
One can verify that
\begin{align}
    \int_{\eta_0}^{\infty} \mathrm{d}\eta_c\,P(\eta_c,\eta_0) = 1\,,
\end{align}
so $P(\eta_c,\eta_0)$ is properly normalized.

\begin{figure}[htpb]
\centering
    \begin{subfigure}[t]{0.48\textwidth}
        \includegraphics[width=\linewidth]{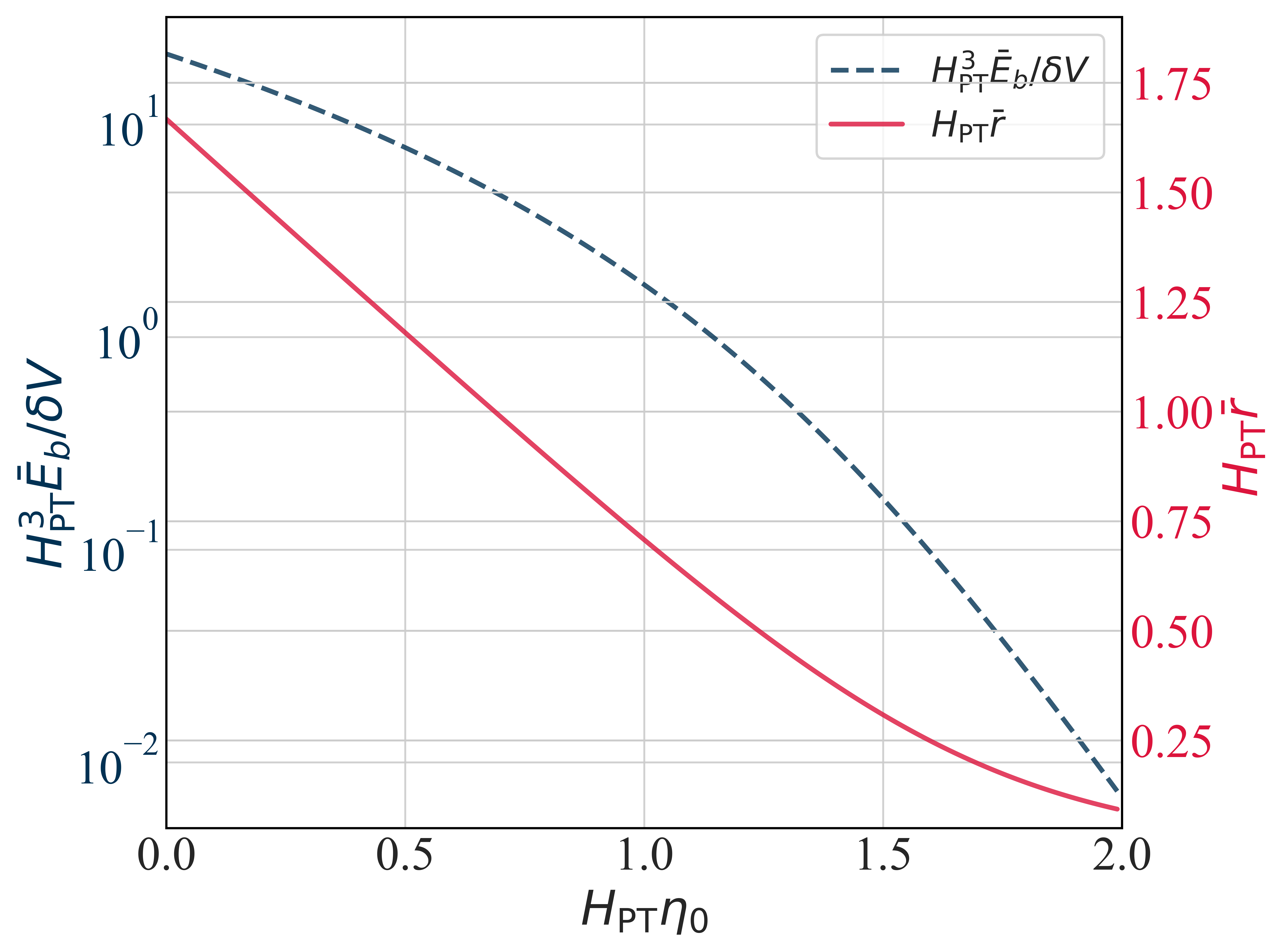}
        \caption{}
        \label{fig:bubble broken time distrib}
    \end{subfigure}
\centering
    \begin{subfigure}[t]{0.48\textwidth}
        \includegraphics[width=\linewidth]{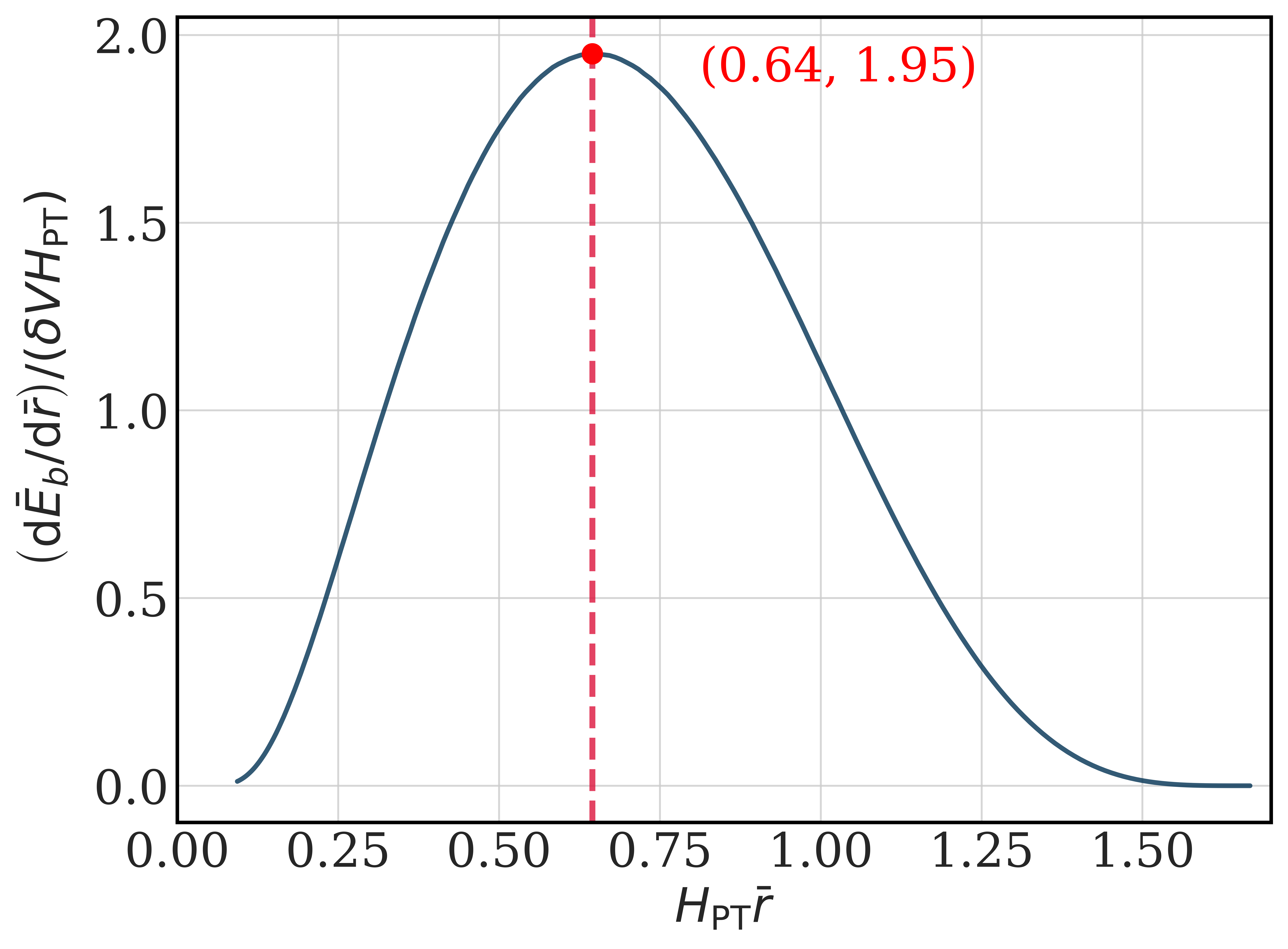}
        \caption{ }
        \label{fig:bubble broken time distrib weight}
    \end{subfigure}
    \caption{In the left panel, the blue and red curves represent the average energy of bubble $\bar{E}_b(\eta_0)$ and the bubble radius $\bar{r}(\eta_0)$, respectively. Both quantities decrease with increasing nucleation time $\eta_0$, as bubbles nucleated later have less time to absorb energy from the false vacuum. In the right panel, the blue curve depicts the derivative $\mathrm{d}\bar{\rho}/\mathrm{d}\bar{r}$. It peaks at $(0.64, 1.95)$, indicating that bubbles carrying the largest energy fraction have a characteristic size of $r \sim 0.64 H_{\mathrm{PT}}^{-1}$. }
    %\label{fig:combined_analysis}
\end{figure}

We now use $P(\eta_c,\eta_0)$ to estimate the contribution of bubbles nucleated at time $\eta_0$ to the GW signal. For a bubble of radius $\eta - \eta_0$, the wall energy scales as $(\eta - \eta_0)^3 a^3(\eta)$. We therefore define the average energy of a bubble nucleated at $\eta_0$ as
\begin{align}
    \bar{E}_b(\eta_0)
    = \delta V \int_{\eta_0}^{\infty} \mathrm{d}\eta_c\,
    (\eta_c - \eta_0)^3 a^3(\eta_c)\,P(\eta_c,\eta_0)\,.
\end{align}
We further define the average comoving radius $\bar{r}(\eta_0)$ of bubbles nucleated at $\eta_0$ through
\begin{align}
    \bar{E}_b(\eta_0)
    = \delta V\,\bar{r}(\eta_0)^3\,
      \big[a\big(\bar{r}(\eta_0) + \eta_0\big)\big]^3 .
\end{align}
The functions $\bar{E}_b(\eta_0)$ and $\bar{r}(\eta_0)$ are shown in Fig.~\ref{fig:bubble broken time distrib}.

Next, we estimate the peak frequency and strength of the GW signal. The comoving number density of bubbles nucleated at $\eta_0$ in a time interval $\mathrm{d}\eta_0$ is
\begin{align}
    \mathrm{d}n(\eta_0)
    = \frac{\Gamma}{\mathcal{V}}\,a^4(\eta_0)\,
      \mathcal{P}_{\rm false}(\eta_0)\,\mathrm{d}\eta_0\,.
\end{align}
Their GW contribution is characterized by $\bar{E}_b(\eta_0)$, while their typical size is given by $\bar{r}(\eta_0)$. We therefore define the radius distribution weighted by the bubble energy,
\begin{align}
    \frac{\mathrm{d}\bar{E}_b(\bar{r})}{\mathrm{d}\bar{r}}
    = \bar{E}_b(\eta_0)\,
      \frac{\mathrm{d}n(\eta_0)}{\mathrm{d}\eta_0}\,
      \left|\frac{\mathrm{d}\eta_0}{\mathrm{d}\bar{r}}\right| .
\end{align}
This distribution is shown in Fig.~\ref{fig:bubble broken time distrib weight}. One finds that the energy-weighted radius distribution peaks at $H_{\mathrm{PT}}\bar{r} \simeq 0.64$. Thus, the peak of the GW signal is expected at a comoving wavenumber of order $\mathcal{O}(1)\,H_{\mathrm{PT}}$.

%Currently, we parameterize the peak momentum as $k_{peak}=c_f H_{\mathrm{PT}}$ with an undetermined constant $c_f$. 

We can further estimate the size of the GW signal.  From Eq.~\eqref{eq:signle bubble tij before collision} we know that The bubble energy momentum tensor is proportional to $\delta V$, which equals to the latent energy density $\Delta \rho$. Then Eq.~\eqref{eq:rho} suggests that $\rho_{\rm GW}\propto {\Delta \rho^2}/{\mpl^2}$. The other scales in $\rho_{\rm GW}$ are all determined by $H_{\mathrm{PT}}$. Using dimension analysis, we obtain that $\rho_{\rm GW}\propto {\Delta \rho^2}/({H_{\mathrm{PT}}^2 \mpl^2})$.
Today's relic GW density spectrum function $\Omega_{\rm GW}(f)$ is determined by comparing the GW energy density to the radiation energy density. Thus, we have 
\bea\label{eq:Omega1}
\Omega_{\rm GW}(f^{\rm peak}) = \frac{\rgw({f^{\rm peak}})}{\rho_{\rm crit}} = c_{\rm GW} \Omega_\gamma \left(\frac{\Delta \rho}{\rho_{\rm tot}}\right)^2 \ ,
\eea
where $c_{\rm GW}$ is a numerical factor and $\rho_{\rm tot}$ is the total energy density of the universe during the phase transition. To derive (\ref{eq:Omega1}) we used the relation $H^2 = 8\pi \gn \rho_{\rm tot}/3$.

In the rest of the paper, we numerically simulate the phase transition and the bubble collision process to calculate $c_f$, $c_{\rm GW}$ and the shape of $\ogw(f)$.

\section{Numerical Simulation of GWs}
\label{sec:numerical}

In this section we present the setup and method of numerical simulation.  In the real case, the bubble walls will achieve extremely large boost when they collide with each other, which is hard to be recovered in lattice simulation.  Also, the comoving size of bubble walls or wall-like objects after collision will decrease since their physical size is fixed. Therefore, although we simulate the full evolution of phase transition field, our simulation can not capture all the properties of the system like the effect of large boost factor on thermalization time scale or the deep IR shape of gravitational waves which requires long range information. 

Therefore, we will focus on the valid regime of numerical simulation and use the semi-analytic method to describe the numerically unreachable parts. 

\subsection{Two bubble collision}\label{sec:two bubble simulation}
For two bubbles colliding on z axis in flat spacetime, the SO(2,1) symmetry ensures that we can use two coordinates $s_1, z$ to describe the system. Here $s_1=\sqrt{t^2-x^2-y^2}$. The EoM  can then be expressed as
\begin{align}
    \frac{\mathrm{d}^2\phi}{\mathrm{d}s_1^2}+\frac{2}{s_1}\frac{\mathrm{d}\phi}{\mathrm{d}s_1}-\frac{\mathrm{d}^2\phi}{\mathrm{d}z^2}+V'(\phi)=0\ .
\end{align}
This implies that $s_1$ serves as the time coordinate during evolution. The initial condition is set by the bounce solution. 

In the simulation, we fix the potential and change the distance between two bubble centers to control the boost factor when collision happens. 
We place one bubble at $z=0$ and the other at $z=r_c$, and track the evolution of the first bubble after the collision. The nucleation time is chosen to be $t_0=0$. Therefore, we have $\rD=\tD$ and $t_c=r_c$ in the simulation. 
The lattice spacing and time step are chosen to match the boost factor. Therefore, the resource scales as $\gamma^4$. We define the boost factor using the critical bubble radius $r_{cr}$ and the collision radius $r_c$, which reads
\begin{align}
    \gamma=\frac{r_c}{r_{cr}}\ .
\end{align}
In Sec.~\ref{sec: flat bubble case}, we derived several analytical predictions for the post-collision evolution of the bubble shell. These include
\begin{align}\label{eq: predictions}
    \phi_{\ell m}^{\rm peak}&\propto r^{-1}(\tD+\rD)^{-1/2}\ ,\\
    \rho_g&\propto  r^{-2}(\tD+\rD)^{-1}\ ,\\
    \Delta r&\propto (\tD+\rD)^1\ ,\\
    r_s&\approx r_c+\tD\ ,
\end{align}
and the total gradient energy per solid angle is conserved. 
%the surface gradient energy density $\sigma_g$ decays as $r^{-2}$, the wall width $\Delta r$ grows as $r^1$, while the field strength $\phi$ decays as $r^{-3/2}$. 
We then use numerical simulations to check these predictions and the effects of nonlinear interactions. 

\begin{figure}[htpb]
\centering
\includegraphics[width=0.99\linewidth]{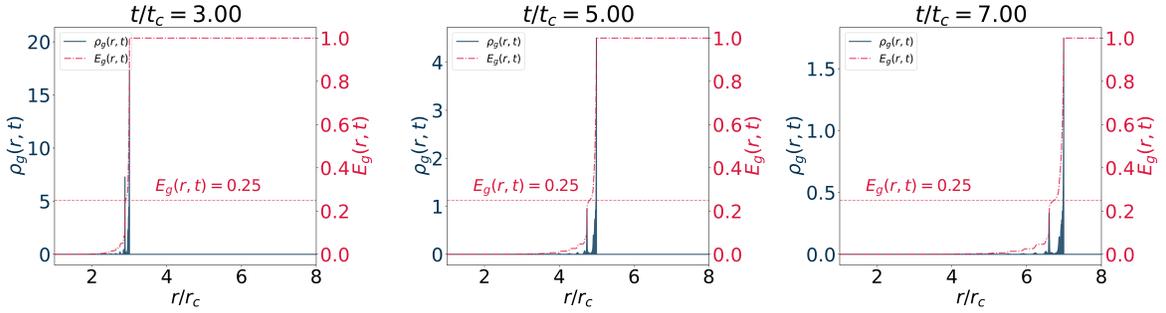}
\caption{The blue and red curves depict the gradient energy distribution and the gradient energy ratio, respectively. In our simulation, two bubbles are initialized at $r=0$ and $r=2r_c$, leading to a subsequent collision at $r=r_c$. We analyze the bubble shell centered at $r=0$ to compute $E_g(r,t)$ and $\rho_g(r,t)$.
The red dotted line indicates the cutoff at $E_g(r,t)=0.25$. This cutoff intersects the gradient energy ratio curve between the first and second shell peaks. As time $t$ evolves, the peak amplitude of the gradient energy decreases; however, the fractional energy contained within the first peak remains constant.}
\label{fig:gradient}
\end{figure}

We simulate a two bubble collision process with a boost factor $\gamma=80$. To track the propagation of energy, we define a gradient energy ratio function per solid angle.
\begin{align}
    E_g(r,t)=\left[E_g(t)\right]^{-1}\int_0^r \rho_g(r',t)r'^2\mathrm{d}r'\ .
\end{align}
Fig.~\ref{fig:gradient} shows the gradient energy ratio function for the bubble shells after collision. The collision initially generates a complex structure with multiple peaks. After the shell propagates several times $r_c$, however, the leading peak — which carries most of the gradient energy — separates from the others. The trailing peaks are left behind, indicating a lower propagation speed. Moreover, the energy fraction contained in the leading peak stabilizes to an approximately constant value after this separation. Consequently, the gradient energy associated with the first peak is also conserved.

%Thus, the main contribution of GWs will attribute to the first bubble shell. 
The next step is to test the predictions in Eq.~\eqref{eq: predictions} against simulations. To isolate the propagating bubble shell from the non-propagating modes generated during collision, we use the first peak in Fig.~\ref{fig:gradient} to calculate the quantities in Eq.~\eqref{eq: predictions}. This peak is shown to carry approximately $75\%$ of the total system's gradient energy.
Therefore, when evaluating $\phi_{\ell m}^{\mathrm{peak}}, \rho_g, \Delta r$, and $r_s$ in the simulation, we set the lower radial bound at the point where $E_g(r,t) = 0.25$, defined as $r_{\mathrm{low}}$. To further suppress numerical fluctuations, we define $\rho_g, \Delta r$, and $r_s$ through spatial integrations, given by
\begin{align}\label{eq: definition of sigma}
    r_s&= \frac{1}{N}\int_{r_{low}}^{\infty}r\left(\frac{\mathrm{d}\phi}{\mathrm{d}r}\right)^2\mathrm{d}r\ ,\\
     (\Delta r)^2&=\frac{1}{N}\int_{r_{low}}^{\infty} (r-r_s)^2\left(\frac{\mathrm{d}\phi}{\mathrm{d}r}\right)^2\mathrm{d}r\ ,\\
     \bar{\rho}_g&=\frac{1}{N}\int_{r_{low}}^{\infty}\rho_g(r,t)\left(\frac{\mathrm{d}\phi}{\mathrm{d}r}\right)^2\mathrm{d}r\ ,
\end{align}
with the normalization factor
\begin{align}\label{eq: definition of r}
    N= \int_{r_{low}}^{\infty} \left(\frac{\mathrm{d}\phi}{\mathrm{d}r}\right)^2\mathrm{d}r\ .
\end{align}
Here, the gradient energy density serves as the weighting function for the integration. Since it shares the same scaling behavior as the quantities being averaged, this choice preserves the scaling laws predicted in Eq.~\eqref{eq: predictions}.
%This quantity provides the information of wall velocity.  The peak amplitude is not easy to keep track on, so instead, we define the average field strength weighted by gradient energy density, which reads
%Since the gradient energy density peaks near the peak of field amplitude, we expect that the average field strength $\bar{\phi}$ also decays as $r^{-3/2}$. 
\begin{figure}[htpb]
\centering
\includegraphics[width=0.9\linewidth]{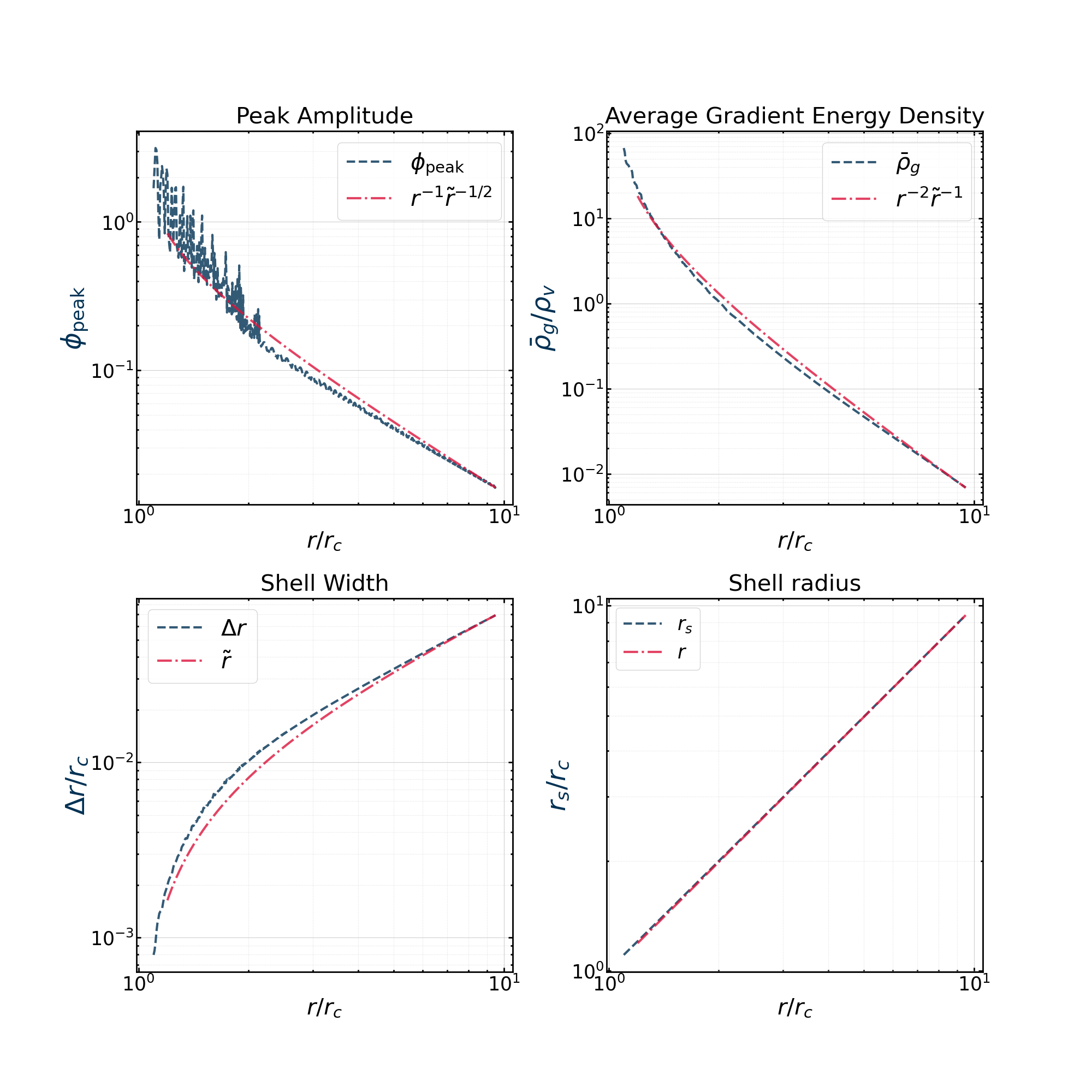}
\caption{The blue curves are the simulation results for the peak amplitude $\phi_{\ell m}^{\rm peak}$, the average gradient energy density $\bar{\rho}_g$, the shell width $\Delta r$, and the shell radius $r_s$ (calculated from Eq.~\eqref{eq: definition of sigma}). The corresponding theoretical scaling predictions are displayed in red. Under the condition $\tD = \rD$ employed in the simulation, the numerical results show good agreement with the theoretical curves. In the simulation, we set the boost factor to $\gamma = 80$. }
\label{fig:predict}
\end{figure}
The simulation data and theoretical predictions are shown in Fig.~\ref{fig:predict}.
%We fit the $r$ dependence of $\sigma_g, r_s,(\Delta r)^2, \bar{\phi}^2$ using a power law ansatz $r^{\alpha}$.
%The result is displayed in Fis.~\ref{fig:predict}.
The deviation between numerical and theoretical results is small. Furthermore, this deviation decreases as $r$ increases, indicating that nonlinear effects decouple when $\rD$ is large. Consequently, the free propagation approximation provides a good description of the  post-collision evolution of bubble shell.

\subsection{GW simulation}
We use a toy model which is similar to the Higgs model with a cubic modification term. The potential is
\begin{align}
    V(\phi)=-\frac{1}{2}m^2\phi^2+\frac{1}{3}\kappa\phi^3+\frac{1}{4}\lambda\phi^4\ ,
\end{align}
where $m^2>0,\kappa>0$. Defining the normalized quantities $x\rightarrow xm, \phi\rightarrow \phi/(m\sqrt{\lambda})$, we thus eliminate $m$ and $\lambda$ from the EoM
\begin{align}
    \frac{\partial^2\phi}{\partial \eta^2}+2\mathcal{H}\frac{\partial \phi}{\partial \eta}-\nabla^2\phi+\left(-\phi+\kappa\phi^2+\phi^3\right)=0\ .
\end{align}
In the following discussion, we will use the normalized quantities. After normalization, although $\lambda$ does not appear in EoM, it still appears as a factor in bounce action and energy-momentum tensor. The bounce action reads
\begin{align}
    S_4=\frac{ 2\pi^2}{\lambda}\int_0^{\infty} \mathrm{d}r r^3\left[\frac{1}{2}\left(\frac{\mathrm{d}\phi_b}{\mathrm{d}r}\right)^2+V(\phi_b)-V(\phi_F)\right]\; ,
    \label{eq: bounce action}
\end{align}
where $\phi_b$ is the bounce solution satisfies Eq.~\eqref{eq: bounce Eom}.  The $\lambda$ parameter in energy-momentum tensor will finally be transferred to GW power spectrum, which will be shown later.  Then the only free parameter in EoM is $\kappa$.  In our simulation, we choose $\kappa=1.7$.  We do not include $\lambda$ in simulation but add it back in the final result $\Omega_{GW}$. 

The basic idea of simulation is to numerically solve the field EoM on a three-dimensional lattice. We build our code on an open source framework AMReX~\cite{DBLP:journals/corr/abs-2009-12009}.   The spatial resolution $\Delta x$ is chosen to be smaller than the width of
bubble walls. Also, the box comoving size should be large enough to include enough bubbles for GW calculation. 
The timestep is chosen to be $\Delta \eta=0.2\Delta x$. 

The bubble generation process can be treated as a Poisson process. In a small interval $\Delta \eta_b$,  the probability that a bubble nucleated inside a volume $(\Delta x)^3$
reads
\begin{equation}
    p=\frac{\Gamma}{V_{\rm phys}}a^4 \Delta\eta_b (\Delta x)^3\; .
    \label{eq: bubble probability}
\end{equation}
For each time interval $\Delta \eta_b$, we use a binomial distribution with probability $p$ to decide whether a bubble generates on each lattice cube. Then we keep those points that their nearby sphere with comoving radius $r_c$ still live in the false vacuum and impose the bounce configuration on these points. Since we use the comoving coordinate, the bounce configuration should be $\phi_b(a\sqrt{\mathbf{x}^2-\Delta \eta_b^2})$. Notice that here we impose a bounce solution with $t>0$. This is because a bounce configuration with $t=0$ is a nearly static solution which does not grow, so we use configuration at $t=a\Delta \eta_b$ to provide it with a small initial momentum. 
The time step evolution is performed by the 4th order symplectic integrator method~\cite{FOREST1990105,Yoshida:1990zz}. 

We also simulate the evolution of $\mathrm{h}_{ij}$ on the lattice using  Eq.~\eqref{eq:GW eom}.
After the phase transition finishes, the Universe is in radiation domination. $\mathrm{h}_{ij}$ then follows the general solution
\begin{align}
    \tilde{\mathrm{h}}_{ij}(k,\eta)=\mathcal{C}_{ij}\sin k\eta+\mathcal{D}_{ij}\cos k\eta\ .
\end{align}
Suppose the simulation stops at $\eta_f$, we then calculate $\mathcal{C},\mathcal{D}$ based on $\mathrm{h}_{ij}(k,\eta_f)$ from simulation, which yields
\begin{align}
    \mathcal{C}_{ij}&=\tilde{\mathrm{h}}_{ij}(k,\eta_f)\sin k\eta_f+\tilde{\mathrm{h}}^{\prime}_{ij}(k,\eta)\frac{\cos k\eta_f}{k}\ ,\\
    \mathcal{D}_{ij}&=\tilde{\mathrm{h}}_{ij}(k,\eta_f)\cos k\eta_f-\tilde{\mathrm{h}}^{\prime}_{ij}(k,\eta)\frac{\sin k\eta_f}{k}\ .
\end{align}
The corresponding GW density power spectrum thus reads
\begin{align}
    \Omega_{\rm GW}=\frac{\Omega_R}{48\pi^2 H_{\mathrm{PT}}^2\lambda^2V}k^5\left[|\mathcal{C}_{ij}^{\rm TT}|^2+|\mathcal{D}_{ij}^{\rm TT}|^2\right]\ .
\end{align}
Here $\lambda$ appears in $\Omega_{\rm GW}$. $\mathcal{C}_{ij}^{\rm TT}$ refers to the transverse-traceless part of $\mathcal{C}_{ij}$ which is defined as $\mathcal{C}_{ij}^{\rm TT}=\Lambda_{ij,kl}\mathcal{C}_{kl}$. The projector $\Lambda_{ij,kl}$ reads
\begin{align}
    \Lambda_{ij,kl}(\mathbf{k})=P_{ik}(\mathbf{k})P_{jl}(\mathbf{k})-\frac{1}{2}P_{ij}(\mathbf{k})P_{kl}(\mathbf{k})\ ,
\end{align}
with
\begin{align}
    P_{ij}(\mathbf{k})=\delta_{ij}-\hat{k}_{\mathrm {eff},i}\hat{k}_{\mathrm {eff},j}\ .
\end{align}
Here $\hat{k}_{\mathrm {eff}}$ refers to the eﬀective momentum which corresponds to the lattice derivative~\cite{Huang:2011gf,Figueroa:2011ye}.

\subsection{Simulation results}
\begin{figure}[htpb]
\centering
    \begin{subfigure}[t]{0.48\textwidth}
        \includegraphics[width=\linewidth]{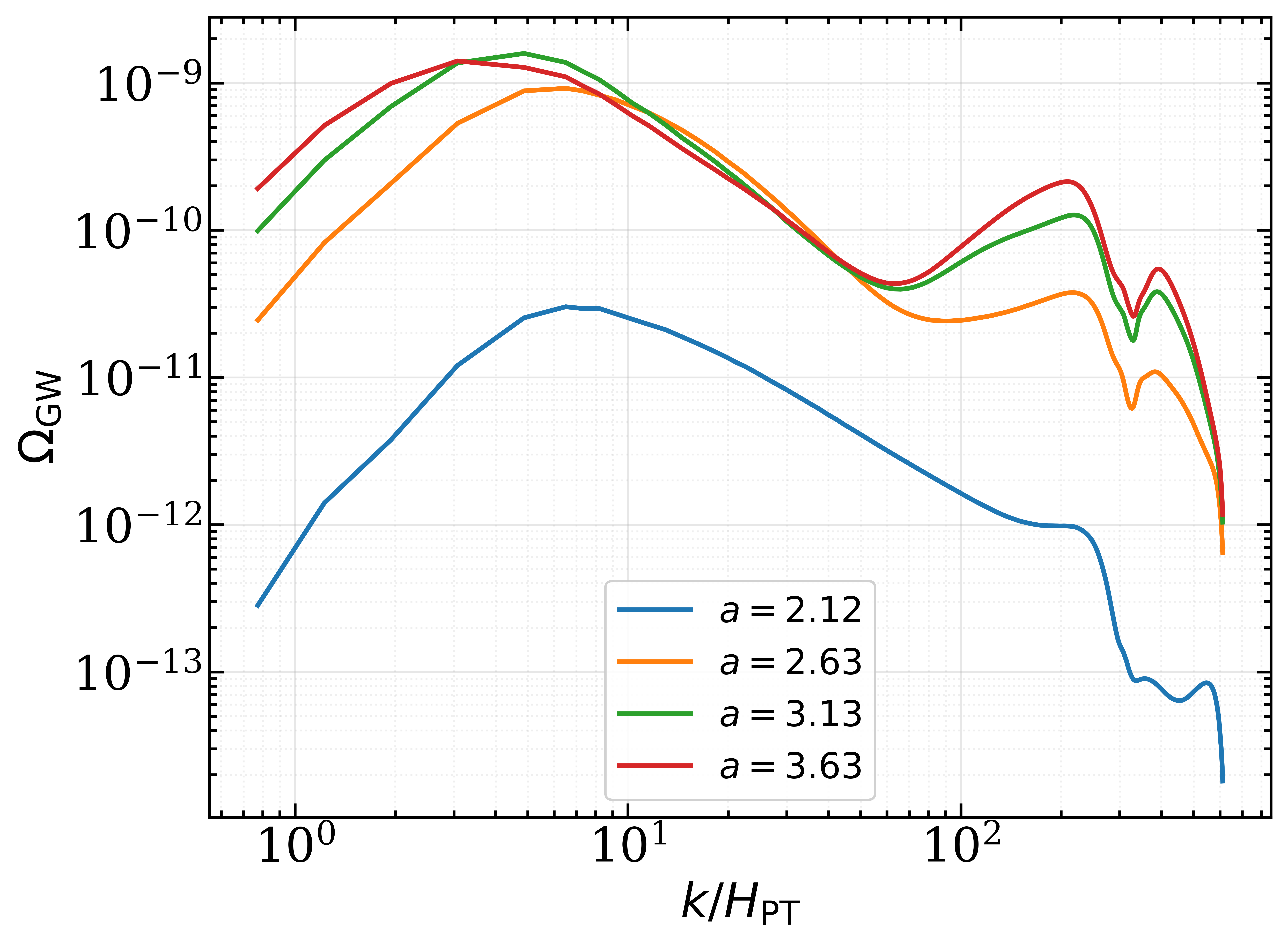}
        \caption{}
        \label{fig:GW 1}
    \end{subfigure}
\centering
    \begin{subfigure}[t]{0.48\textwidth}
        \includegraphics[width=\linewidth]{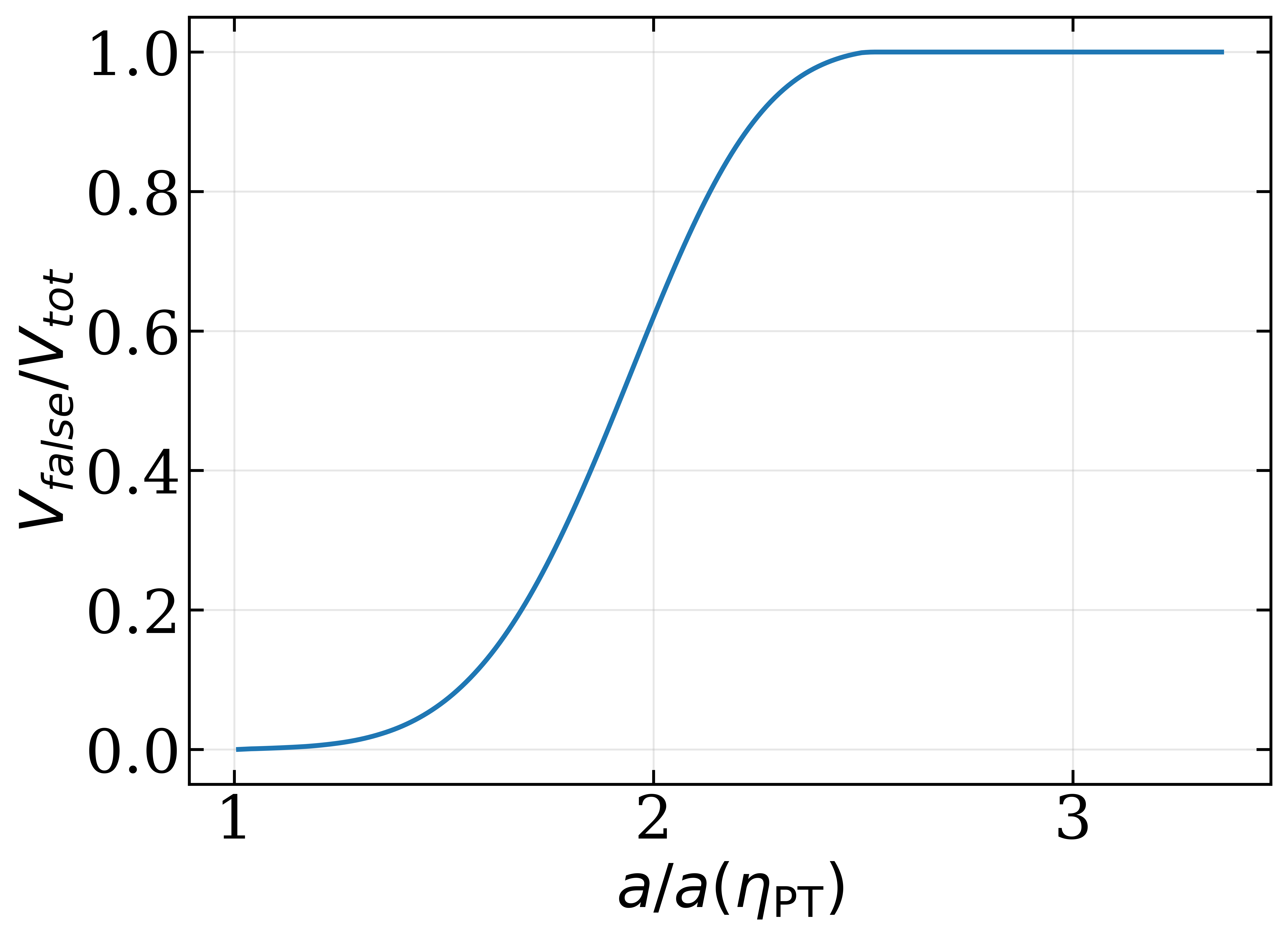}
        \caption{ }
        \label{fig:fv fraction}
    \end{subfigure}
    \caption{(a) Gravitational wave power spectrum evaluated at different times. The spectrum initially grows and then stabilizes to a constant value as the phase transition approaches completion.(b) True vacuum fraction from the simulation. For $a/a_{\mathrm{PT}} > 2.5$, the true vacuum occupies most of the volume, indicating that the phase transition has effectively concluded.}
\end{figure}

In Sec.~\ref{sec:fv fraction}, we obtain a constraint for the energy density of the dark sector using $\Delta N_{\mathrm{eff}}$, which gives
\begin{align}
    \frac{\Delta \rho}{3 H_{\mathrm{PT}}^2\mpl^2}\leq 0.02\ .
\end{align}
The vacuum energy density in our simulation is set according to this constraint. The resulting gravitational wave power spectrum is shown in Fig.~\ref{fig:GW 1}. The spectrum continues to grow until $a/a_{\mathrm{PT}} \approx 3$. In the simulation, the limited boost factor prevents the emergence of the long-term bubble shell evolution; consequently, the accumulation of gravitational waves ceases approximately one e-fold after the phase transition concludes.

\begin{figure}[htpb]
\centering
\includegraphics[width=0.8\linewidth]{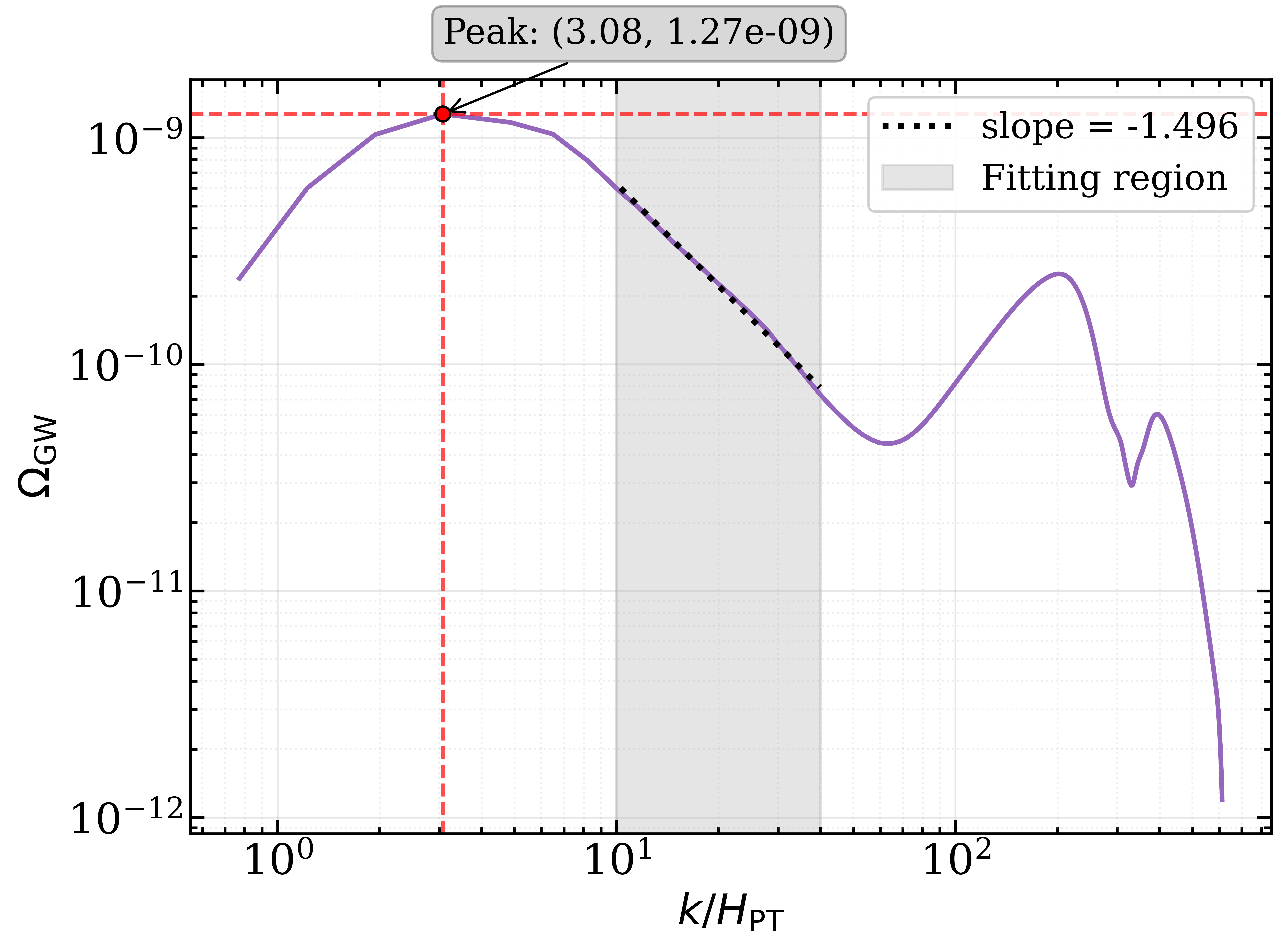}
\caption{The gravitational wave energy density spectrum, $\Omega_{\mathrm{GW}}$, is evaluated at a scale factor of $a/a(\eta_{\mathrm{PT}})=3.26$, chosen to the time when the gravitational wave power spectrum has stabilized. Our simulation adopts parameters $\Omega_\gamma=10^{-5}$ and $\Delta\rho/\rho=0.02$, yielding Eq.~\eqref{eq:fitting}.
The spectrum exhibits a power-law behavior $\propto k^{-1.50}$ within the gray-shaded region. In the deep UV regime, a double-peak structure emerges at wavenumbers close to the mass scale of the phase transition field. However, we note that this particular feature is likely a simulation artifact; in a realistic cosmological context, the relevant Hubble scale at this epoch is expected to be far removed from the mass scale of the dark sector field, rendering such a prominent double peak physically implausible. }
\label{fig:GW 2}
\end{figure}

Fig.~\ref{fig:fv fraction} shows the true vacuum fraction from the simulation. The phase transition completes around $a/a_{\mathrm{PT}} \approx 2.5$, indicating that the gravitational wave source does not vanish immediately after bubble collisions have ended.
The best fitting for GW strength is 
\begin{align}\label{eq:fitting}
    \Omega^{\rm peak}_{\mathrm{GW}}=1.27\times 10^{-4}\Omega_{\gamma} \left(\frac{\Delta \rho}{0.02\rho_{\rm tot}}\right)^2\ ,
\end{align}
with the spectral peak located at
\begin{align}
    k_{peak}=3.08 H_{\mathrm{PT}}\ .
\end{align}
It is difficult to obtain the precise shape of infrared part of $\Omega_{\mathrm{GW}}$ from simulation, but the UV behavior is clear in simulation. The envelope approximation prediction of UV shape is $k^{-1}$, while the bulk flow model prediction is $k^{-2}$. From Fig.~\ref{fig:GW 2} the UV slope is about $k^{-1.5}$, which is between the prediction of the bulk flow model and the envelope model in flat spacetime. 
%This provides collateral evidence that in dark phase transition, the evolution of bubble walls after collision should be described by bulk flow model instead of envelope approximation. 
The deviation of the UV slope from the bulk flow model in flat spacetime could result from the RD background.

\section{PBH Abundance}
\label{sec:PBH}

First-order phase transition naturally shares properties of supercooled phase transitions since the nucleation rate is a constant and the majority of bubbles nucleate when $\Gamma/\mathcal{V}\sim H^4$. Therefore, the time scale of this kind of phase transition can easily reach Hubble time scale. During phase transition process, suppose a metastable vacuum island survives for several efolds, the vacuum energy will dominate the inside of this island and trigger it to inflate. If the physical size of the vacuum island exceeds its Schwarzschild radius while the over-density is large enough, it may collapse into a PBH.

In this section, we will calculate the probability of finding such a false vacuum island at a given time which will be useful for estimating the number density of PBHs. We assume that during the phase transition, the universe is in radiation domination, and the Standard Model sector dominates the evolution of the universe.

\subsection{Vacuum islands distribution}
The nucleation and expansion of true vacuum bubbles cut the false vacuum into finite-sized domains. The exact structures of these false vacuum domains(FVDs) are complicated, which means it is hard to describe their distribution in detail. But we can still apply some simplifications to their shapes and obtain a rough description of FVDs. We will use a method similar to~\cite{Kodama:1982sf,Lewicki:2023ioy}. 

In section \ref{sec:fv fraction} we calculate the fraction of false vacuum through Eq.~\eqref{eq:false vacuum retio}. 
It can be generalized to the following expression. 
\begin{align}
    {\cal P}_{\rm false}(r, t) = \exp\left[ - \frac{\Gamma}{\mathcal{V}} \int_0^t \D t' \frac{4\pi}{3} \left(r+R(t,t')\right)^3 a^3(t') \right]  \ ,
\label{eq:false vacuum general}
\end{align}
${\cal P}_{\rm false}(r, t)$ refers to the probability of finding a point whose nearby spherical region with comoving radius $r$ lies inside the false vacuum at time $t$. 
Notice that it is not equivalent to finding a FVD with radius $r$ since this probability does not ensure that the region outside this domain has been occupied by true vacuum. 
\begin{figure}[htpb]
\centering
\includegraphics[width=0.6\linewidth]{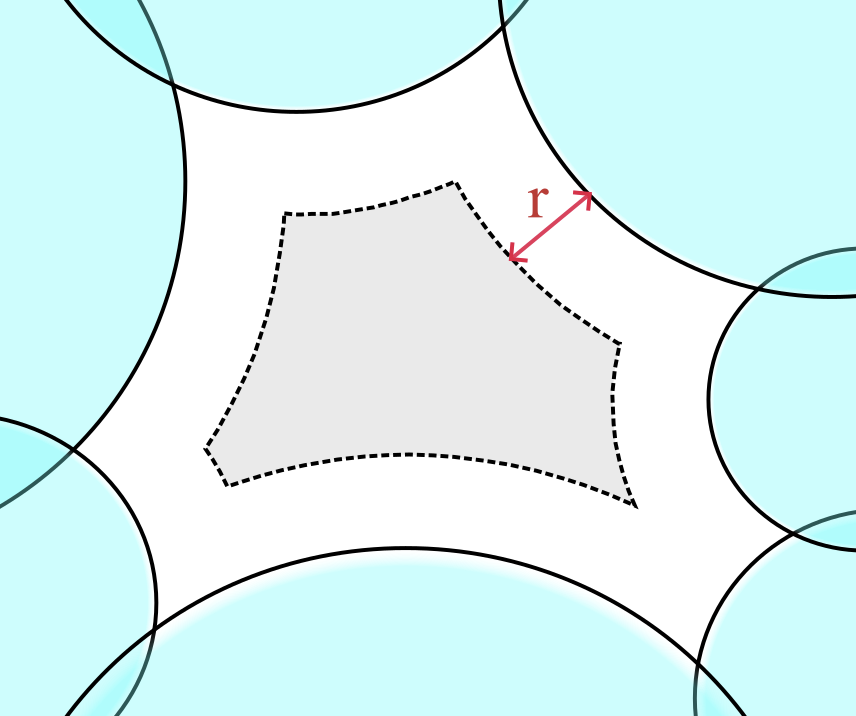}
\caption{The blue regions are occupied by bubbles which represent the true vacuum part. The probability function ${\cal P}_{\rm false}(r, t)$ describes the fraction of the gray region whose boundary has a distance of $r$ from the true-false vacuum boundary. }
\label{fig: false vacuum domain}
\end{figure}
As shown in Fig.~\ref{fig: false vacuum domain}, ${\cal P}_{\rm false}(r, t)$ refers to the region whose distance to the bubble wall equals to $r$. When $r=0$, ${\cal P}_{\rm false}(r, t)$ then recovers the result of false vacuum portion ${\cal P}_{\rm false}(t)$. We define
\begin{align}\label{eq: log distribution function}
    \mathcal{A}(r,t) = -\ln {\cal P}_{\rm false}(r, t)=\frac{\Gamma}{\mathcal{V}} \int_0^t \D t' \frac{4\pi}{3} \left(r+R(t,t')\right)^3 a^3(t')\ .
\end{align}
Although ${\cal P}_{\rm false}(r, t)$ does not directly give us the distribution of vacuum islands, we can still obtain some information about the size of vacuum islands from it. In general, the shape of false vacuum can be arbitrary and nonspherical. 

We will first try to use a simplified model to describe this system. Suppose the false islands can be approximated as a set of ellipsoids with different radius. The number density distribution function of these ellipsoids can be represented as $n(t, r_1,r_2,r_3)$. We use the convention that $r_1\geq r_2\geq r_3$ (which means when we write down $n(t,r_1,r_2,r_3)$, we treat it as $n(t, r_1,r_2,r_3)\theta(r_1-r_2)\theta(r_2-r_3)$). Then ${\cal P}_{\rm false}(r, t)$ can be expressed by $n$, which yields
\begin{align}\label{eq: ellipsoid expansion}
    {\cal P}_{\rm false}(r, t)=\frac{4\pi}{3}\iiint\mathrm{d}r_1 \mathrm{d}r_2\mathrm{d}r_3 n(t, r_1,r_2,r_3)(r_1-r)(r_2-r)(r_3-r)\theta(r_3-r)\ .
\end{align}
Here $\theta$ is the step function. Applying $\frac{\partial^4}{\partial r^4}$ on both sides of Eq.~\eqref{eq: ellipsoid expansion}, we obtain
\begin{align}\label{eq: ellipsoid expansion partial}
    \frac{\partial^4}{\partial r^4}{\cal P}_{\rm false}(r, t)=8\pi\iint\mathrm{d}r_1 \mathrm{d}r_2n(t, r_1,r_2,r)\ .
\end{align}
The right side of Eq.~\eqref{eq: ellipsoid expansion partial} refers to the total number density of islands with their semi-minor axis larger than $r$. Since the probability for a vacuum island to survive decreases exponentially when we increase $r$, we expect that the majority of ellipsoid islands with $r_3=r$ should be sphere like in large $r$ limit, which means
\begin{align}\label{eq: sphere distribution}
    n(t,r_1,r_2,r)\rightarrow \mathcal{N}(t,r)\delta(r_1-r)\delta(r_2-r)\ ,
\end{align}
when $r\rightarrow \infty$. Here $\mathcal{N}(t,r)$ describes a distribution function of sphere islands. Inserting Eq.~\eqref{eq: sphere distribution} into Eq.~\eqref{eq: ellipsoid expansion partial}, yields
\begin{align}\label{eq: number density of sphereical islands}
    \mathcal{N}(t,r)=\frac{1}{8\pi}\frac{\partial^4}{\partial r^4}{\cal P}_{\rm false}(r, t)\ .
\end{align}
Therefore, when we consider large vacuum islands, we can use Eq.~\eqref{eq: number density of sphereical islands} to estimate the number density of vacuum islands.

\subsection{PBH threshold}
\label{sec: threshold}

In our discussion, we will employ the Schwarzschild criterion to determine whether PBHs form or not. The same criterion has also been applied in \cite{Baker:2021sno,Jung:2021mku,Lewicki:2023ioy,Flores:2024lng}. Since there is an event horizon once the condition is satisfied, this condition does not depend on whether the vacuum island possesses spherical symmetry. The mass of a spherical island is~\cite{Blau:1986cw}
\begin{equation}
  M=\frac{4\pi r^3}{3}\rho_{\rm false}+4\pi\sigma r^2 \sqrt{1+\left(\frac{d r}{d \tau}\right)^2-H_{\rm false}^2r^2}-8\pi^2 G_{\rm N} \sigma^2 r^3\ ,
\end{equation}
where $\tau$ is the proper time of the bubble wall. In this equation, the first term is the energy of false vacuum, the second term is the energy of the bubble wall, and the third term is the self-gravitational energy of bubble wall. For a Hubble size island, $r\simeq H_{\rm false}^{-1}$, the rest energy of bubble wall is smaller than the vacuum energy,
\begin{equation}
    \frac{r^2\sigma }{r^3\rho_{\rm false}}\sim\frac{m}{\mpl}\ll1\ .
\end{equation}
On the other hand, due to the non-sphericity of the bubble wall, its energy cannot enter the event horizon all at once during black hole formation. So the contribution of the bubble wall to the black hole mass is minor. In this case, we neglect the energy of the bubble wall. The criterion means there is a radius $r$ satisfies
\begin{equation}\label{eq: S criterion}
	\frac{2\gn M(r)}{r}=1\ ,
\end{equation}
with
\begin{equation}
	M(r)\equiv \int_{0}^{r}\rho(r')4\pi r'^2 dr'\ .
\end{equation}
Since we only consider the vacuum energy of the false vacuum island $\rho(r)=\rho_{\rm false}$, the condition is equivalent to the size of the island is larger than its Hubble radius,
\begin{equation}\label{eq: radius}
	r_{\rm false}\ge H_{\rm false}^{-1}\ .
\end{equation}
Therefore, an observer in the exterior region would perceive the formation of a black hole, while an observer in the interior would observe an inflating universe. This interior region, which is causally disconnected from our universe, actually corresponds to a baby universe~\cite{Harada:2004pe,Kopp:2010sh,Carr:2014pga}. This implies that, when the energy of the bubble wall is neglected, applying the Schwarzschild criterion yields type-II PBHs. Such a scenario can also arise from the collapse of sufficiently large curvature perturbations generated during inflation~\cite{Uehara:2024yyp,Shimada:2024eec,Escriva:2025eqc,Escriva:2025rja,Uehara:2025idq}, as well as in other ultra-slow cosmological phase transitions~\cite{Jinno:2023vnr,Zhong:2025xwm}.

It should be noted that when applying the Schwarzschild criterion, if the energy density of the vacuum island is not significantly greater than that of the exterior, what is actually identified is a cosmological horizon rather than a black hole event horizon. In other words, there is no external region of a black hole in this case. Therefore, an additional condition is required that outside the horizon of the vacuum island, there must exist a cosmological horizon of the external FRW spacetime. And our universe corresponds to the region between these two horizons. This condition implies the existence of another solution $r>H_{\rm false}^{-1}$ for Eq.~\eqref{eq: S criterion}. Assuming the average energy density for the region $r>H_{\rm false}^{-1}$ is $\rho_{\rm out}$, we have
\begin{equation}
	M(r)\simeq \frac{4\pi}{3}\left[H_{\rm false}^{-3}\rho_{\rm false}+\left(r^3-H_{\rm false}^{-3}\right)\rho_{\rm out}\right]\ .
\end{equation}
Eq.~\eqref{eq: S criterion} becomes
\begin{equation}
	1-\left(\frac{H_{\rm out}}{H_{\rm false}}\right)^2+\left(\frac{H_{\rm out}}{H_{\rm false}}\right)^2\left(rH_{\rm false}\right)^3-rH_{\rm false}=0\ ,
\end{equation}
with $H_{\rm out}=\sqrt{\frac{8\pi \gn\rho_{\rm out}}{3}}$. The solutions of this equation are
\begin{equation}
	r_0=H_{\rm false}^{-1}\ ,\ \ \ \ r_\pm=\frac{\pm\sqrt{4H_{\rm false}^2-3H_{\rm out}^2}-H_{\rm out}}{2H_{\rm out}}H_{\rm false}^{-1}\ .
\end{equation} 
Since $H_{\rm false}>H_{\rm out}$, we can see $r_-$ is negative and our condition requires $r_+>r_0$, this is
\begin{equation}
	H_{\rm false}\ge\sqrt{3}H_{\rm out}\ .
\end{equation}
Then the criterion of density contrast is
\begin{equation}\label{eq: contrast}
	\delta_{\rm c}\equiv\frac{\rho_{\rm false}-\rho_{\rm out}}{\rho_{\rm out}}=2\ ,
\end{equation}
which is much larger than the criterion of the collapse of curvature perturbations from inflation. Because in our case, the space is flat and there is no space curvature that aids the collapse~\cite{Sasaki:2018dmp,Flores:2024lng}. So a vacuum island can form a PBH only when Eq.~\eqref{eq: radius} and Eq.~\eqref{eq: contrast} are satisfied. If the energy density of true vacuum is $\rho_{\rm true}$, PBH formed after $\rho_{\rm false}\ge 3\rho_{\rm ture}$. However, in practice, we don't need to apply this criterion directly. Because if a vacuum island is entirely surrounded by radiation, as long as no new bubbles nucleate inside it, its density contrast will always exceed 2 later. Here, we adopt the percolation time $t_p$ as the reference time for counting the number density of vacuum islands when the true vacuum becomes connected and the false vacuum turns into isolated regions.

Furthermore, if the energy of the bubble wall is not neglected, its contribution may allow a vacuum island with a size smaller than $H_{\rm false}^{-1}$ to form a PBH as well. Such a PBH would be a conventional Type-I PBH, which does not lead to the creation of a baby universe. However, in the absence of spherical symmetry, it is difficult for these bubble walls to collapse entirely within the Schwarzschild radius without undergoing collisions. Consequently, it is challenging to accurately estimate the PBH production rate in this scenario without numerical simulations. We therefore disregard this case for the moment and defer it to a more detailed future investigation.

\subsection{PBH relic abundance}

Using Eq.~\eqref{eq: background eom}, we can also get $\rho_{\rm true}$ and $\rho_{\rm false}$ separately with the vacuum energy density $\rho_v$ satisfying
\begin{equation}
    \rho_{v}=\left\{
    \begin{matrix}
        &\mathcal{P}_{\rm false}( t)\Delta \rho\ ,&{\rm for\ true\ vacuum}\\
        &\Delta \rho\ .&{\rm for\ false\ vacuum}
    \end{matrix}\right.
\end{equation}

\begin{figure}[htpb]
\centering
\includegraphics[width=0.7\linewidth]{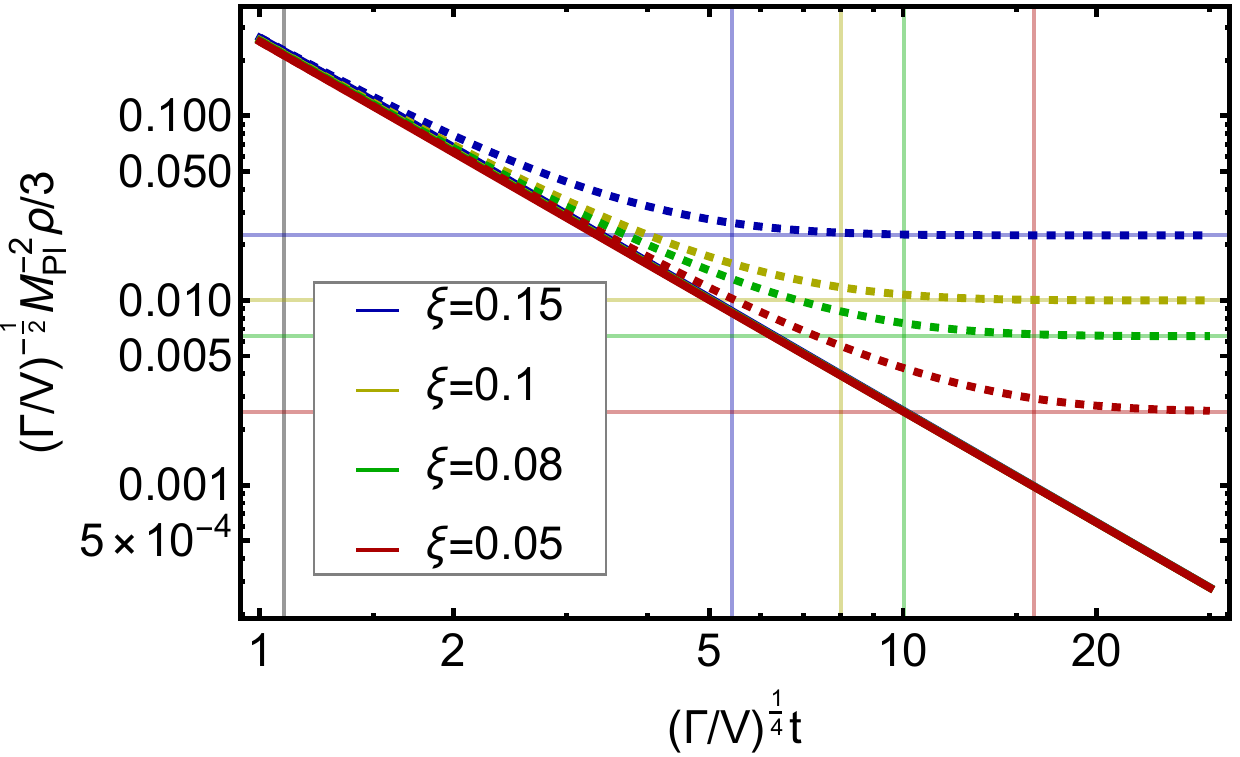}
\caption{The temporal evolution of the energy density for different $\xi$. The cases $\xi=0.05,\ 0.08,\ 0.1,\ 0.15$ are represented by red, green, yellow and blue, respectively. The solid line denotes the energy of true vacuum while the dashed line denotes the energy of false vacuum. The percolation time $t_p$ is shown in gray vertical line and the PBH production time $t_{\rm PBH}$ is shown in colored vertical lines. And the colored horizontal lines represent the latent energy density $\Delta \rho$. }
\label{fig: vn}
\end{figure}

\begin{figure}[htpb]
\centering
\includegraphics[width=0.6\linewidth]{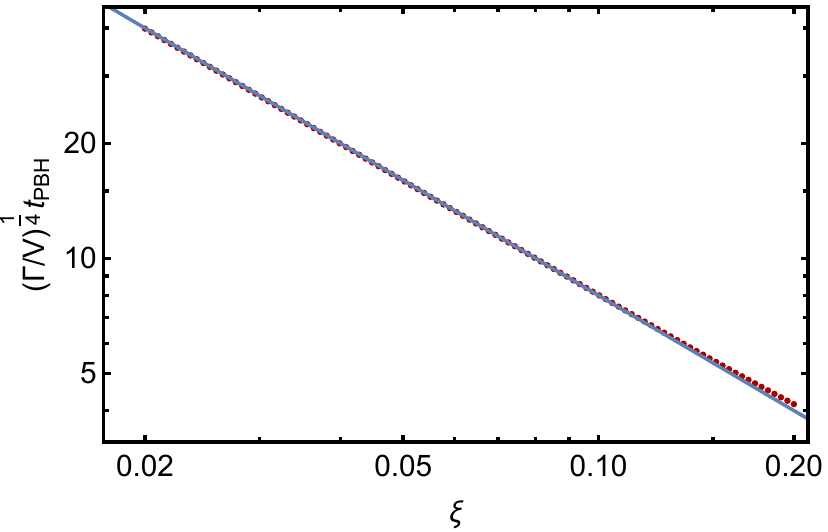}
\caption{The relation between $\xi$ and PBH production time $t_{\rm PBH}$. The red dotted line is the result from simulation while the blue solid line is the fitting result $\left(\Gamma/{\cal V}\right)^{\frac{1}{4}}t_{\rm PBH}\simeq0.8\xi^{-1}$.}
\label{fig: dc}
\end{figure}

In Fig.~\ref{fig: vn}, we have depicted the temporal evolution of the energy density in the true vacuum and false vacuum regions for different $\xi$. The evolution of the universe is unaffected by the phase transition since the solid curves are superimposed. It can be seen that around $t_{\rm PBH}$ when $\rho_{\rm false}(t_{\rm PBH})=3\rho_{\rm true}(t_{\rm PBH})$ the energy density of false vacuum can be estimated by the latent energy density $\Delta \rho$, so the physics size of the false vacuum island is
\begin{equation}
    H_{\rm false}^{-1}\simeq\left(\frac{\Gamma}{\cal V}\right)^{-\frac{1}{4}}\xi^{-1}\ .
\end{equation}
The physical size of those islands at $t_p$ is
\begin{equation}
    r_{\rm island}=\frac{a(t_p)}{a(t_{\rm PBH})}H_{\rm false}^{-1} +a(t_p)R(t_p,t_{\rm PBH})=\left(\frac{t_p}{t_{\rm PBH}}\right)^{\frac{1}{2}}\left(\frac{\Gamma}{\cal V}\right)^{-\frac{1}{4}}\xi^{-1}+2\left(\sqrt{t_pt_{\rm PBH}}-t_p\right)\ .
\end{equation}
The PBH production time $t_{\rm PBH}$ for different $\xi$ is shown in Fig.~\ref{fig: dc}. It's can be observed that
\begin{equation}
    t_{\rm PBH}\simeq0.8\left(\frac{\Gamma}{\cal V}\right)^{-\frac{1}{4}}\xi^{-1}\ .
\end{equation}

\begin{figure}[htpb]
\centering
\includegraphics[width=0.6\linewidth]{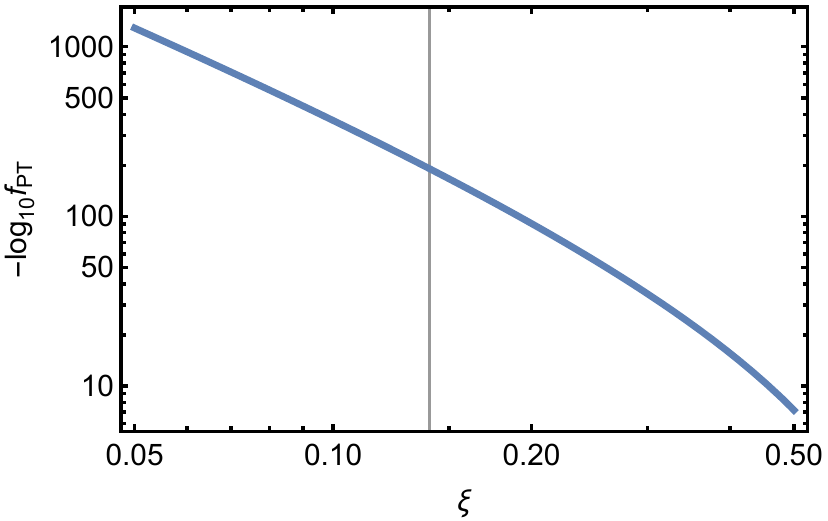}
\caption{The relation between $\xi$ and PBH fraction $\frac{a_p}{a_{\rm eq}}f_{\rm PBH}$.}
\label{fig: logf}
\end{figure}

During the radiation dominated era, the probability that a region is still in the false vacuum can be calculated as
\begin{equation}
    \mathcal{P}_{\rm false}\left(r_{\rm phys},t\right)=\exp\left[-\frac{\Gamma}{\cal V}\frac{4\pi}{3}\left(\frac{2}{5}r_{\rm phys}^3t+\frac{2}{5}r_{\rm phys}^2t^2+\frac{8}{35}r_{\rm phys}t^3+\frac{2}{35}t^4\right)\right]\ ,
\end{equation}
where $r_{\rm phys}=a(t)r$ is the physical radius.
Then the number of islands per comoving volume within a physical radius region $\left(r_{\rm phys},r_{\rm phys}+d r_{\rm phys}\right)$ is
\begin{equation}
    \mathcal{N}\left(r_{\rm phys},t\right)d r_{\rm phys}=\frac{1}{8\pi}a^3(t)\frac{\partial^4}{\partial r_{\rm phys}^4}\mathcal{P}_{\rm false}\left(r_{\rm phys},t\right)dr_{\rm phys}\ .
\end{equation}
The comoving number density of PBH is
\begin{equation}
    n_{\rm PBH}=\int_{r_{\rm island}}^\infty \mathcal{N}\left(r,t_p\right)dr=-\frac{a_p^3}{8\pi}\frac{\partial^3}{\partial r^3}\mathcal{P}_{\rm false}\left(r_{\rm island},t_p\right)\ ,
\end{equation}
then
\begin{equation}
    f_{\rm PBH}=\frac{n_{\rm PBH}M_{\rm PBH}}{a_{\rm today}^33\mpl^2H_{\rm today}^2\Omega_{\rm DM}}\simeq\frac{a_{\rm eq}}{a_p}\left(\frac{t_p}{t_{\rm PBH}}\right)^2\frac{8\pi n_{\rm PBH}}{a_p^3H_{\rm false}^3}\ ,
\end{equation}
where $a_p=a(t_p)$, $a_{\rm eq}$ is the scale factor at radiation-matter equality and the mass of PBH is 
\begin{equation}
    M_{\rm PBH}\simeq\frac{1}{2\gn H_{\rm false}}\simeq 6.7\times10^{-11} \xi^{-1}\left(\frac{(\Gamma/{\cal V})^{1/4}}{{\rm eV}}\right)^{-1}M_\odot\ .
\end{equation}
The factor $\frac{a_{\rm eq}}{a_p}$ is determined by $\frac{\Gamma}{\cal V}$,
\begin{equation}
    \frac{a_{\rm eq}}{a_p}\simeq\left(\frac{H_p}{H_{\rm eq}}\right)^{\frac{1}{2}}\simeq4.4\times 10^{13}\left(\frac{(\Gamma/{\cal V})^{1/4}}{{\rm eV}}\right)^{\frac{1}{2}}
\end{equation}
while the energy fraction of PBH at the percolation time
\begin{equation}
    f_{\rm PT}\equiv \frac{a_p}{a_{\rm eq}}f_{\rm PBH}\ ,
\end{equation}
is only determined by $\xi$. The dependence is shown in Fig.~\ref{fig: logf}. It can be seen that the $\Delta N_{\rm eff}$ constraint is stronger than the PBH constraint. For cases satisfied the $\Delta N_{\rm eff}$ constraint $\xi\le 0.14$ the PBH is not likely to be formed. 

However, if the DS  energy fraction decreases after phase transition, for example, the DS can release energy to SM sector, the constraint on $\xi$ can be relaxed. Then for $(\Gamma/{\cal V})^{1/4}\simeq 1\ \mathrm{eV}$, $\xi\sim 0.4$, the PBH from DS phase transition has a mass of the order of asteroid mass and can explain all the dark matter.

\section{Summary and Outlook}
\label{sec:summary}
In this work, we study the first order phase transition of a dark sector which lies in a pure metastable vacuum and has no non-gravitational interactions with the SM. The nucleation rate of this kind of remains a constant and the phase transition occurs when $\Gamma/\mathcal{V}\approx H^4$. The average radius of bubbles can thus reach Hubble size and there is no suppressed factor like $H/\beta$ in GW amplitude. 

In a dark sector phase transition, the boost factor of bubble walls before collision can reach extremely high values and persists even after collision due to the absence of interactions with a background plasma. These shell-like profiles continue to propagate, maintaining a thin structure for an extended period.

In Minkowski spacetime, the shell radius grows as $r^1$, while its gradient energy density falls as $r^{-3}$. Consequently, the total gradient energy of the spherical shell is conserved.
In a radiation-dominated universe, Hubble friction causes the shell's energy density to decay more rapidly and its width to grow faster. The total energy decays as $\eta^{-2}$, and the field strength decays as $\eta^{-7/2}$ in the long-term evolution. This rapid decay suggests that nonlinear interactions become negligible quickly.
In the radiation-dominated case, the width of the bubble shell becomes comparable to its radius when the cutoff momentum of the pre-collision bubble wall becomes non-relativistic. This condition estimates when the bubble shell disappears, neglecting nonlinear interactions.

The long-term evolution of the bubble shell after collision modifies the IR behavior of the gravitational wave power spectrum. It introduces an extra contribution proportional to $k^3\log^2(k/H_{\mathrm{PT}})$ to the IR slope. This modification is typically absent in numerical simulations because the achievable boost factor is limited; consequently, the dilution of bubble shells occurs much earlier than in the physical scenario.

Nevertheless, numerical simulations can still accurately capture the peak and UV behavior of the power spectrum. From our simulation, the best-fit peak amplitude of the GW energy density spectrum is
\bea
\Omega^{\rm peak}_{\mathrm{GW}} \approx 0.3~  \Omega_{\gamma} \left(\frac{\Delta\rho}{\rho_{\rm tot}}\right)^2 \ ,
\eea
with the peak located at 
\bea
k_{\mathrm{peak}} \approx 3.1 H_{\mathrm{PT}} \ .
\eea
The UV slope is approximately $k^{-1.5}$, which is  different from the shape predicted by the bulk flow model in Minkowski spacetime.
A detailed understanding of the bubble shell dynamics immediately following collision is needed to accurately describe the UV slope of the GW spectrum; this investigation is left for future work.

For DS phase transitions, a constant decay rate implies that, for a given transition strength, more false-vacuum regions survive compared to a thermal phase transition. This enhances the formation of Type-II PBHs. However, if the latent heat of the DS phase transition is fully converted into dark radiation, the stringent bounds from $\Delta N_{\rm eff}$ strongly suppress PBH production, rendering it negligible. By contrast, if non-gravitational interactions between the DS and the Standard Model are present, these constraints can be relaxed, allowing for a significantly larger PBH abundance.

The decay of a DS metastable vacuum can be used to generate dark matter~\cite{Freese:2023fcr}. However, in such scenarios it is difficult to probe the nature of dark matter through anything other than its gravitational effects.

\appendix

\section{The effective EoM for bubble expansion before collision}
\label{sec:AppendixA}
In this appendix, we derive the effective EoM for the
wall before the collision. The procedure follows that in~\cite{Lewicki:2023ioy}. The full action of $\phi$ reads
\begin{align}\label{eq:bubble action}
    S=\int\mathrm{d}\eta\mathrm{d}x^3 a^4\left[\frac{1}{2a^2}(\partial_\eta\phi)^2-\frac{1}{2a^2}(\nabla\phi)^2-\frac{1}{2}V(\phi)\right]\ .
\end{align}
We then insert ansatz Eq.~\eqref{eq:bubble wall ansatz} into the action. The temporal derivative terms yields
\begin{align}\label{eq:bubble eft action}
    \partial_\eta \phi=-a\gamma v_w\partial_{\tilde{r}}\phi_s+\partial_\eta(a\gamma)(r-r_s)\partial_{\tilde{r}}\phi_s\ .
\end{align}
Here we define the wall-frame coordinate $ \tilde r \equiv a(\eta)\,\gamma(\eta)\,[r - r_s(\eta)] $.
Since the wall width is small, $\partial_{\tilde{r}}\phi_s$ is non zero when $r-r_s\sim (\gamma m)^{-1}$. On the other hand, $\partial_\eta(a\gamma)\sim a\gamma/\eta$. Therefore, the second term in Eq.~\eqref{eq:bubble eft action} is suppressed by a factor $1/\gamma m \eta\sim \gamma ^{-2}$ compared to the first term. The spatial derivative terms read
\begin{align}\label{eq:bubble eft action r}
    \nabla \phi=a\gamma\partial_{\tilde{r}}\phi_s\ .
\end{align}
Combined with Eq.~\eqref{eq:bubble eft action}, the derivative parts in the action can be expressed as
\begin{align}
    \frac{1}{2}(\partial_\eta\phi)^2-\frac{1}{2}(\nabla\phi)^2&=\frac{1}{2}\left[a^2\gamma^2 v_w^2-2\tilde{r}v_w\partial_\eta(a\gamma)+ \left(\partial_\eta(a\gamma)(r-r_s)\right)^2-a^2\gamma^2\right](\partial_{\tilde{r}}\phi_s)^2\ ,\nonumber\\
    &=\frac{1}{2}\left[-a^2-2\tilde{r}v_w\partial_\eta(a\gamma)+ \left(\partial_\eta(a\gamma)(r-r_s)\right)^2\right](\partial_{\tilde{r}}\phi_s)^2\ ,\nonumber\\
    &\approx \frac{1}{2}\left[-a^2-2\tilde{r}v_w\partial_\eta(a\gamma)\right](\partial_{\tilde{r}}\phi_s)^2 \ .
\end{align}
Here we use the relation $\gamma^2(1-v_w^2)=1$. In the third line we neglect the term $\left(\partial_\eta(a\gamma)(r-r_s)\right)^2$ because it is suppressed by $\gamma^{-2}$ compared to the first two terms based on the previous discussion. 
Then, we can complete the spatial integration in Eq.~\eqref{eq:bubble action}, which yields
\begin{align}
    \int\mathrm{d}x^3&\left[\frac{1}{2a^2}(\partial_\eta\phi)^2-\frac{1}{2a^2}(\nabla\phi)^2-\frac{1}{2}V(\phi)\right]\nonumber\\
    &=\frac{2\pi}{a^2}\int r^2\mathrm{d}r\left[-a^2-2\tilde{r}v_w\partial_\eta(a\gamma)\right](\partial_{\tilde{r}}\phi_s)^2+\frac{2}{3}\pi r_s^3\delta V \nonumber\ , \\
    &\approx  \frac{2\pi r_s^2}{a^3\gamma}\int \mathrm{d}\tilde{r}\left[-a^2-2\tilde{r}v_w\partial_\eta(a\gamma)\right](\partial_{\tilde{r}}\phi_s)^2+\frac{2}{3}\pi r_s^3\delta V \nonumber\ .
\end{align}
Using the definition of wall tension $\sigma$ and the condition 
\begin{align}\label{eq:rw}
    \int \tilde{r}\mathrm{d}\tilde{r}(\partial_{\tilde{r}}\phi_s)^2=0\ ,
\end{align}
since $\phi_s$ is a kink-like configuration. Eq.\eqref{eq:rw} can also be treated as the definition of $r_s(\eta)$. Now we can obtain the effective action for bubble wall, which reads
\begin{align}
    S_{wall}=\int\mathrm{d}\eta \left[a^4 \frac{2}{3}\pi r_s^3\delta V-2\pi a^3 r_s^2 \sigma \sqrt{1-(\partial_\eta r_s)^2}\right]\ .
\end{align}
The EoM of the wall thus yields
\begin{align}
    \partial_\eta\left(\gamma v_w\right)+3 \mathcal{H}\left(\gamma v_w\right)+\frac{2 \gamma}{r_s}=a \frac{\delta V}{\sigma}\ .
\end{align}

\section*{Acknowledgments}

This work is supported in part by the National Key R\&D Program of China under Grants Nos. 2021YFC2203100 and 2017YFA0402204, the NSFC under Grant Nos. 12475107 and 12525506, and the Tsinghua University Dushi Program.

\bibliography{refs}
\bibliographystyle{JHEP}

\end{document}